\documentclass[5p,authoryear,12p]{elsarticle}

\usepackage{amssymb}

\journal{Icarus}

\def\apj{ApJ}%
\def\aap{A\&A}%
\def\icarus{Icarus}%
\def\jgr{J.~Geophys.~Res.}%
\begin{document}

\begin{frontmatter}



\title{A 1D microphysical cloud model for Earth, and Earth-like exoplanets\\ Liquid water and water ice clouds in the convective troposphere}


\author[mpia]{Andras Zsom\corref{cor1}}
\author[mpia,cfa]{Lisa Kaltenegger}
\author[canada]{Colin Goldblatt}

\address[mpia]{Max-Planck-Institute f\"ur Astronomie, K\"onigstuhl 17, D-69117 Heidelberg, Germany.}
\address[cfa]{Harvard-Smithsonian Center for Astrophysics, 60 Garden Street, Cambridge, MA 02138, USA}
\address[canada]{School of Earth and Ocean Sciences, University of Victoria, BC, Canada}

\cortext[cor1]{Corresponding author. Currently working at the Department of Earth, Atmospheric and Planetary Sciences, Massachusetts Institute of Technology, Cambridge, MA 02139. Email: \texttt{zsom@mpia.de}, or \texttt{zsom@mit.edu}}

\begin{abstract}
One significant difference between the atmospheres of stars and exoplanets is the presence of condensed particles (clouds or hazes) in the atmosphere of the latter. In current 1D models clouds and hazes are treated in an approximate way by raising the surface albedo, or adopting measured Earth cloud properties. The former method introduces errors to the modeled spectra of the exoplanet, as clouds shield the lower atmosphere and thus modify the spectral features. The latter method works only for an exact Earth-analog, but it is challenging to extend to other planets.

The main goal of this paper is to develop a self-consistent microphysical cloud model for 1D atmospheric codes, which can reproduce some observed properties of Earth, such as the average albedo, surface temperature, and global energy budget. The cloud model is designed to be computationally efficient, simple to implement, and applicable for a wide range of atmospheric parameters for planets in the habitable zone.

We use a 1D, cloud-free, radiative-convective, and photochemical equilibrium code originally developed by Kasting, Pavlov, Segura, and collaborators as basis for our cloudy atmosphere model. The cloud model is based on models used by the meteorology community for Earth's clouds. The free parameters of the model are the relative humidity and number density of condensation nuclei, and the precipitation efficiency. In a 1D model, the cloud coverage cannot be self-consistently determined, thus we treat it as a free parameter.

We apply this model to Earth (aerosol number density 100 cm$^{-3}$, relative humidity 77 \%, liquid cloud fraction 40\%, and ice cloud fraction 25\%) and find that a precipitation efficiency of 0.8 is needed to reproduce the albedo, average surface temperature and global energy budget of Earth. We perform simulations to determine how the albedo and the climate of a planet is influenced by the free parameters of the cloud model. We find that the planetary climate is most sensitive to changes in the liquid water cloud fraction and precipitation efficiency. 

The advantage of our cloud model is that the cloud height and the droplet sizes are self-consistently calculated, both of which influence the climate and albedo of exoplanets.

\end{abstract}


\end{frontmatter}


\section{Introduction}
Clouds on exoplanets influence the temperature-pressure profile (T-P profile) of the atmosphere owing to their high opacity and therefore influence the borders of the Habitable Zone \citep[see e.g.,][]{Forget1997, Selsis2007, Kitzmann2010}. They also influence the detectable spectra by reducing the observable atmospheric features \citep[see e.g.,][]{Kaltenegger2007,Kitzmann2011}. Clouds are also a source of uncertainty in problems such as the Faint Young Sun Paradox \citep[see e.g.,][]{Sagan1972, Goldblatt2011}. Earth, Venus, and to a small extent, Mars all have clouds, therefore we expect that exoplanets in the Habitable Zone (HZ) also harbor clouds in their atmospheres.

The climatic effects of clouds can be mimicked by increasing the surface albedo \citep[see][and subsequent work]{Kasting1984}. Although this method can capture the climatic effects of clouds and reproduce the observed pressure-temperature structure of Earth, the measured global mean annual energy budget of Earth cannot be reproduced \citep{Goldblatt2011}. Furthermore, the effects of clouds on the observed spectra is neglected by this approach. To address this issue, the measured cloud properties of Earth can be used in models assuming that the cloud properties do not change for different exoplanets \citep{Kitzmann2010}. An alternative approach is to take the cloud heights and droplet sizes from observations, and fit the cloud fractions and liquid water paths in each cloud layer to match the global energy budget (GEB) of Earth \citep{Goldblatt2011}. Such models adopt multiple cloud layers in agreement with observations \citep{Rossow2005, Warren2007}. A parameterized cloud model is difficult to adopt for exoplanets which are not the exact duplicate of Earth because the number of cloud layers, and other cloud properties (such as the condensing material, height of the layers, droplet sizes, cloud fraction) are all unconstrained parameters in the general case. 

Due to the large amount of observational and laboratory data available for Earth, detailed theoretical cloud models have been developed by the meteorological community and are available for 1D vertical \citep[see e.g.,][]{Srivastava1967}, 2D \citep[see e.g.,][]{Fan2007, Shima2009}, and 3D general circulation models \citep[see e.g.,][]{Stephens2005}. We do not use these models because they are either too Earth-centric and study one specific cloud type on a local scale (e.g., cumulus or cirrus clouds only), and/or they are too computationally expensive and thus not suitable for parameter space exploration. Instead we develop a new 1D model using basic and well-understood physics also employed in these models.

We develop a new 1D microphysical cloud model which is suitable for a wide range of planetary and atmospheric parameters for exoplanets in the HZ. Our cloud model has five free parameters, and does not add a significant computational overhead to existing 1D atmosphere models. 

The structure of the paper is as follows. We describe the numerical methods and codes used for the atmosphere modeling in Sect. \ref{sec:nummet}. The optical properties of liquid and solid water droplets and our microphysical cloud model is discussed in Sect. \ref{sec:cloudm}. We give the initial conditions of the simulations and perform convergence tests in Sect. \ref{sec:tests}. We perform four sets of simulations to study the climatic effects of the five free parameters of the cloud model in Sect. \ref{sec:res}. Finally, we discuss and summarize our results in Sects. \ref{sec:disc} and \ref{sec:sum}.

\section{Numerical methods}
\label{sec:nummet}
\subsection{The 1D radiative-convective and photochemical code}
\label{sec:rad-conv}
We use a radiative-convective climate model \citep{Kasting1993, Pavlov2000} that is coupled to a photochemical model \citep{Pavlov2002, Segura2003, Segura2005, Segura2010, Kaltenegger2011a}. The radiative-convective model determines the vertical temperature and pressure profiles of the atmosphere in radiative equilibrium for a fixed chemical composition. The photochemistry code determines the vertical mixing ratio profiles of different chemical species in chemical equilibrium for a fixed temperature and pressure profile. The two modules are used iteratively until both radiative and chemical equilibria are simultaneously reached. 

The photochemistry code was originally developed by \cite{Kasting1985}. It solves for 55 chemical species that are linked by 220 reactions and it uses a reverse-Euler method to solve the chemical network \citep{Segura2010}. 

The climate model consists of two parts. The first part calculates the transfer of stellar radiation in the visible and near-infrared. The second part calculates the transfer of thermal infrared radiation of the planet. Knowing the difference in the incoming and outgoing fluxes and the heat capacity of the atmospheric gases, we calculate the temperature change of each layer:
\begin{equation}
\frac{dT}{dt} = \frac{g}{c_p} \frac{dF}{dP},
\label{eq:dTdt}
\end{equation}
where $T$ is the temperature of the layer, $dt$ is the time step, $g$ is the gravitational constant, $c_p$ is the total heat capacity of the layer that depends on the composition of the atmosphere, $dF$ is the net flux propagating through layer (difference between incoming and outgoing fluxes), $dP$ is the pressure difference at the top and bottom of the layer. If the temperature gradient ($dT/dz$ -- lapse rate) in a given layer is larger than the adiabatic lapse rate, convective adjustment is performed.

\subsection{Changes in the 1D code}
\label{sec:changes}

To implement clouds into the 1D model, we need high vertical resolution (small $dz$), because clouds only occupy a small vertical portion of the atmosphere. Typically the height of the atmospheres we consider are about 70 km with a cloud layer of a few hundred meters, although deep convective clouds around the equator can be several kilometers thick \citep{Mace2009}. Therefore the number of vertical layers is a free parameter in our code and convergence tests are used to determine the appropriate number of layers to be used (see Sect. \ref{sec:conv_test}). 

We decrease the time step of the calculations and increase the number of iterations compared to the cloud-free version code due to the high resolution of our grid (see Sect. \ref{sec:inicond} for more details). We update the convergence test of the original code to examine the relative change of temperature and mixing ratio of important greenhouse gases between two successive calls in \emph{each} layers individually.

The radiative transfer module of the original code treated scattering on hydrocarbon hazes in the atmosphere \citep{Domagoldman2008, Haqq-Misra2008}. We produced new optical data tables for liquid and icy water droplets. The new values were calculated for 80 different log-spaced droplet radius values between 0.1 and 100 microns.

\section{The cloud model}
\label{sec:cloudm}


\subsection{Optical properties of cloud droplets}
We assume that the liquid droplets are spherical and deformation due to aerodynamical flows around the droplets can be neglected. Ices have a crystal structure that depends on the environment they form in, and the structure influences their optical properties \citep{Liou1993}. These shape effects are crucial in Earth atmospheric sciences where observational data in good quality is abundant, but such effects are secondary in exoplanet atmosphere and paleo-climate problems. Therefore we neglect these shape effects and use the equivalent sphere assumption.

\begin{figure*}
\centering
  \includegraphics[width=0.49\textwidth]{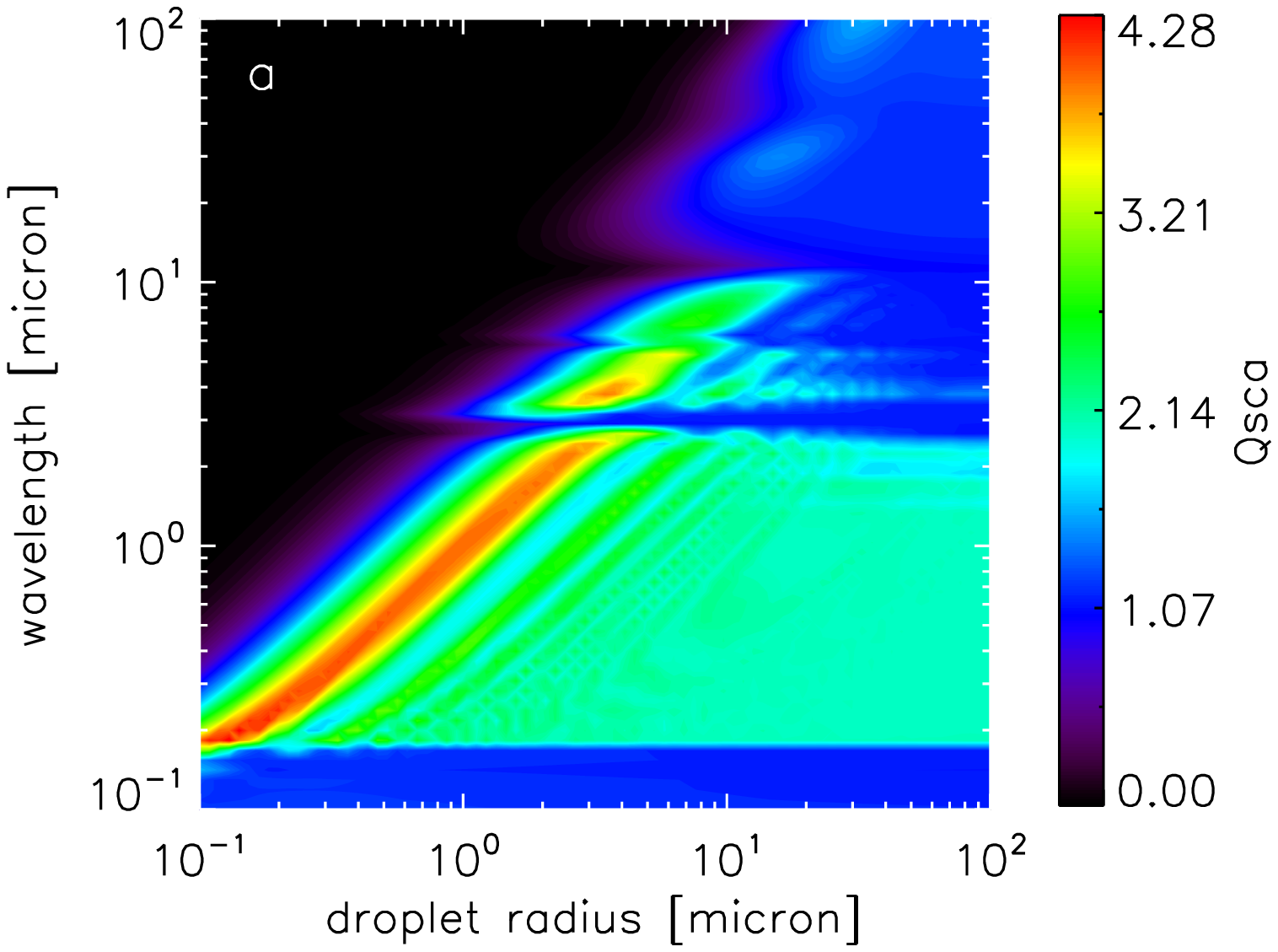}
  \includegraphics[width=0.49\textwidth]{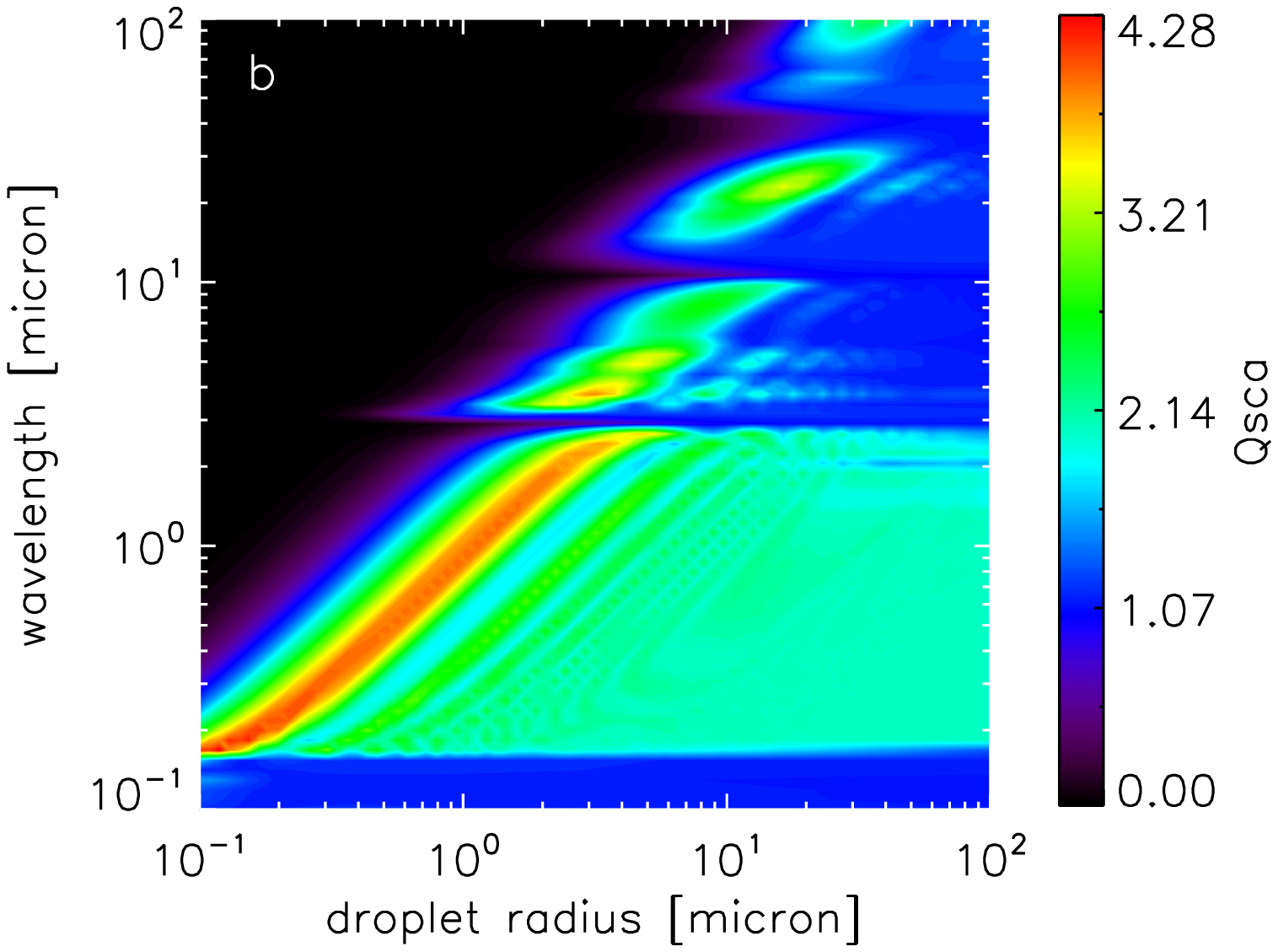}
  \caption{Scattering coefficient of liquid water droplets as a function of droplet radius and wavelength (\ref{fig:sca}a), same for water ice (\ref{fig:sca}b).}
  \label{fig:sca}
\end{figure*}

\begin{figure*}
\centering
  \includegraphics[width=0.49\textwidth]{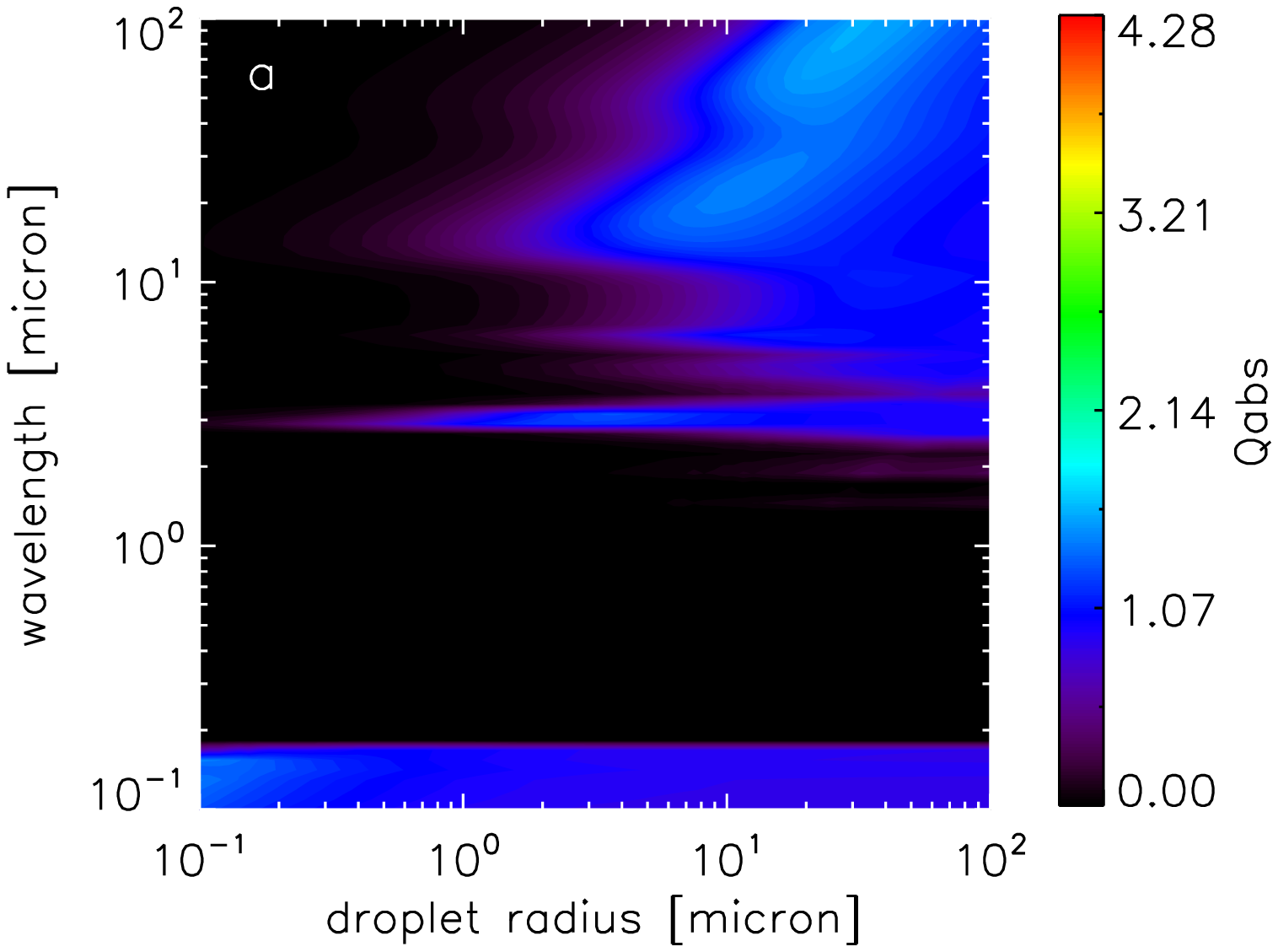}
  \includegraphics[width=0.49\textwidth]{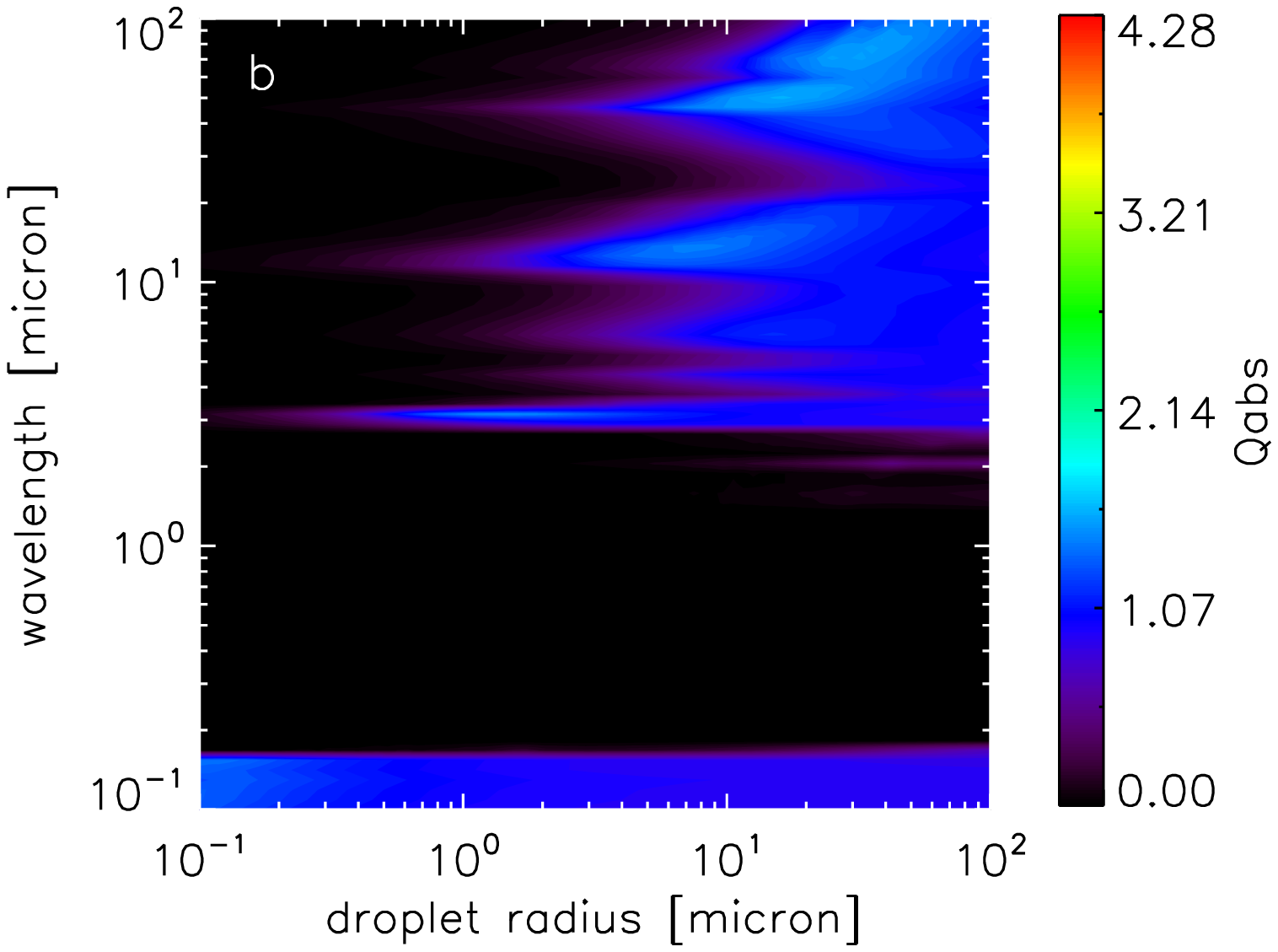}
  \caption{Absorption coefficient of liquid water droplets as a function of droplet radius and wavelength (\ref{fig:abs}a), same for water ice (\ref{fig:abs}b).}
  \label{fig:abs}
\end{figure*}

\begin{figure*}
\centering
  \includegraphics[width=0.49\textwidth]{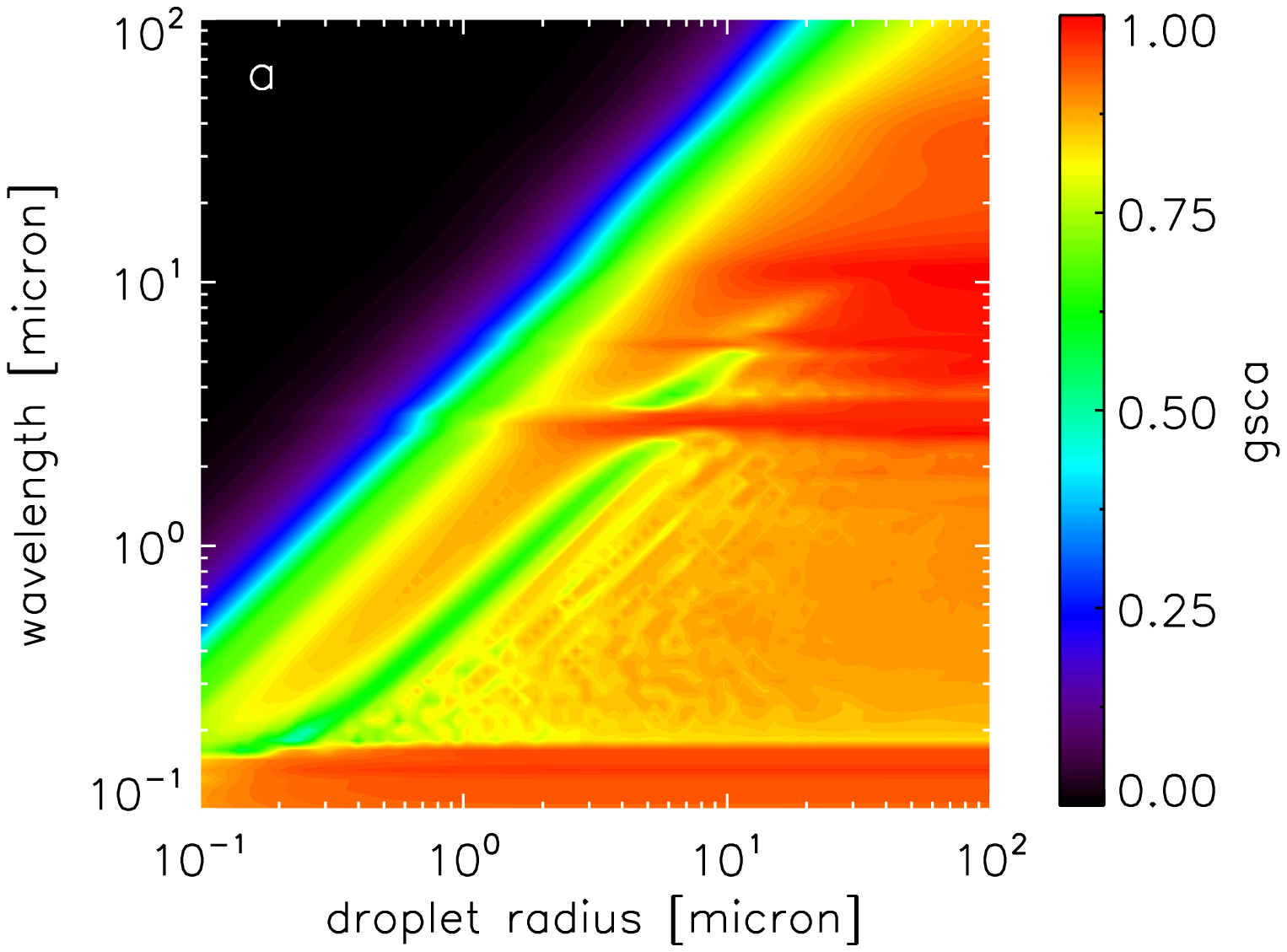}
  \includegraphics[width=0.49\textwidth]{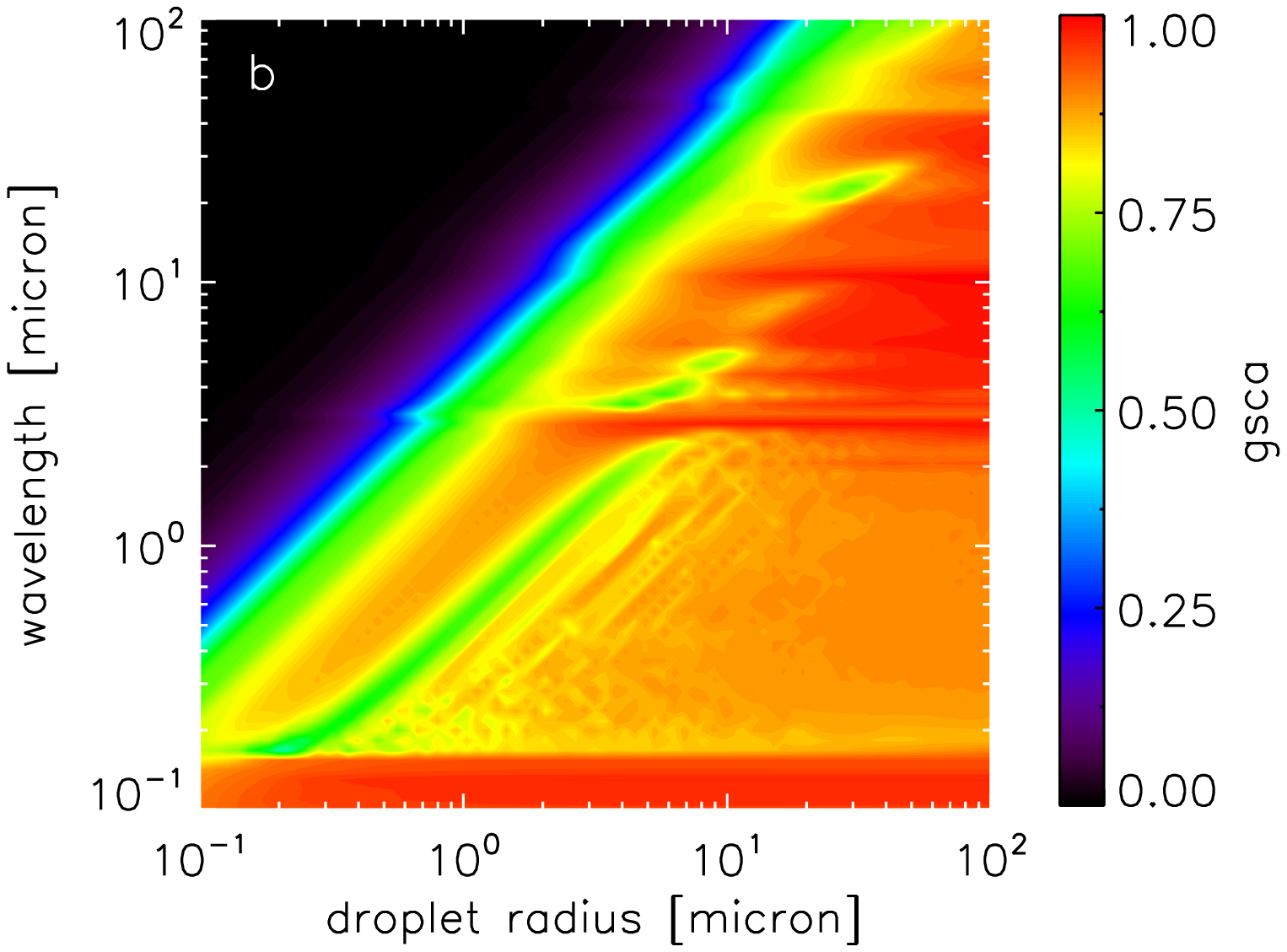}
  \caption{Asymmetry parameter of liquid water droplets as a function of droplet radius and wavelength (\ref{fig:gsca}a), same for water ice (\ref{fig:gsca}b).}
  \label{fig:gsca}
\end{figure*}

The complex refractive index value of liquid water \citep{Segelstein1981} and water ice \citep{Warren2008} is used to calculate the optical properties of droplets as a function of radius and wavelength using Mie calculations \citep{Bohren1983}. The absorption ($Q_{abs}$), scattering coefficients ($Q_{sca}$), and the asymmetry parameter ($g_{sca}$) of the droplets are shown on Figs. \ref{fig:sca}, \ref{fig:abs}, and \ref{fig:gsca}, respectively. These coefficients are converted into the cross-section area of a single droplet:
\begin{equation}
\sigma = Q (2\pi r),
\end{equation}
where $\sigma$ is the (absorption or scattering) cross-section, $Q$ is the corresponding coefficient, and $r$ is the droplet radius. The asymmetry parameter describes whether the droplet is an isotrop scatterer ($g_{sca}=0$; light is scattered to all directions uniformly), forward scatterer ($g_{sca}=1$; analog of a perfect lens) or backward scatterer ($g_{sca}=-1$; analog of a perfect mirror). All of the above describe quantities are angle independent for spherical particles.



The diagonals of Figs. \ref{fig:sca}, and \ref{fig:abs} show that the absorption and scattering coefficients are larger than unity, if the photon's wavelength is similar to the droplet's radius. In this case, the cross-section area of the droplet is several times its geometrical area due to a resonance effect between the light and the droplet.

Droplets between 0.2 and 3 microns are strong scatterers but hardly absorb light. This wavelength range coincides with the peak of the solar spectral energy distribution (SED). However, if the star's SED is more pronounced at wavelengths shorter than 0.2 microns, or longer than 3 microns, the water droplets become more efficient absorbers. The amount of light scattered or absorbed by the clouds thus depends on the spectral type of the star as well as on the size distribution of droplets.

The most significant difference between liquid water and water ice particles is that ice particles are strong scatterers, but less efficient absorbers in the thermal IR for sizes around 10 micron (see Fig. \ref{fig:sca}b). Due to this difference, ice particles scatter the outgoing thermal radiation more efficiently than liquid droplets.

\subsection{Definitions and basic considerations}
\label{sec:basic}
We adopt and use terminology from the meteorology community in the next sections, therefore we clarify some definitions and explain some concepts. The saturation water vapor pressure is described by the Clausius-Clapeyron equation and is a function of temperature. The pressure dependence of the saturation vapor pressure can be neglected under atmospheric conditions. Relative humidity is the ratio of partial water vapor pressure to the saturation vapor pressure. If the relative humidity is larger than unity, the air is in a super-saturated state. The amount of excess water vapor pressure relative to the saturation level is called the level of super-saturation. The saturation pressure relative to liquid water and water ice is calculated according to \cite{Rogers1989}. If the atmosphere is saturated relative liquid water, it is super-saturated relative to water ice.

The process of condensation (or nucleation) can be heterogenous or homogeneous \citep{Rogers1989}. Homogenous nucleation means that the water vapor molecules collide with each other and spontaneously form stable water droplets. This process is not observed in the atmosphere for two reasons. 1) Homogenous nucleation requires extreme levels of super-saturation (several times the saturation vapor pressure), however the super-saturation levels observed in the atmosphere are only a few percentage at maximum. 2) Small, sub-micron sized aerosols are always present in the atmosphere. The water molecules preferentially collide with these aerosols and are deposited on their surfaces. This process is called heterogenous nucleation. Therefore, heterogenous nucleation removes the excess water vapor before homogenous nucleation could be initiated.

These aerosols or condensation nuclei can be naturally produced by erupting volcanoes (ash particles), bursting bubbles of salty water (as a bubble on the ocean or sea bursts, the tiny droplets are transported away from the surface and dried by the wind to form tiny salt crystals), dust particles from the land, the agglomeration of various sulfate, nitrate, ammonium or organic chemicals, or bacteria and spores can serve as condensation nuclei \citep{Seinfeld2006}. Aerosols are also produced by anthropogenic sources on Earth. The number density and size distribution of these aerosols can be measured on Earth \citep[see e.g.,][]{Meszaros1974a, Hoppel1990}, but their properties are unknown on exoplanets. Therefore we assume that some amount of abioticly produced aerosols are present in all atmospheres such that heterogenous nucleation can be initiated.

The formation of cloud droplets can be initiated by increasing the relative humidity above the saturation level in two ways. 1) Adding moisture to the air, or radiatively cooling it without displacing the parcel. This is how fog on the ground forms. 2) Increase the altitude of the air parcel. In such an environment, the air parcel adiabaticly expands and cools until the saturation level is reached after which condensation starts. 

Cloud properties on Earth are studied in detail using satellite measurements \citep{Rossow2005, Mace2009}. Large-scale features of Earth cloud patterns in the tropics include the inter-tropical convergence clouds, the rising branch of Hadley circulation with typically deep convective clouds, and the cloud-free descending branch of the Hadley circulation. In contrast, mid-latitudes are dominated by low level clouds.

Clouds can form as a result of convection during which the height of clouds can be larger than their horizontal extent (cumulus). Opposed to this, clouds can be horizontally extended (stratus) or form thin threads (cirrus). Stratus and cumulus clouds can be formed at low, and mid-altitudes, and they are called alto-stratus or alto-cumulus in the latter case. Cirrus clouds are ice clouds thus only appear at high altitudes favorable for ice formation. Cirrus clouds can also form via convection (cirrocumulus) or form an extended horizontal layer (cirrostratus). There are also towering clouds, which occupy more than one height ranges (cumulonimbus and nimbostratus clouds).

There are still many open questions related to cloud physics, and we mention only a few of these here. It is an active field of research how the relative importance of radiative, gravity wave, or precipitation processes influence convective clouds \citep{Stevens2005}. It is not fully understood how the formation and life time of cirrus clouds is influenced by different nucleation and radiative processes \citep{Jensen1996, Spichtinger2010}. Perhaps low level stratus clouds are the least well understood because their properties (such as the height of the cloud base) change by small and slowly varying perturbations in radiative forcing, precipitation, and evaporation of falling droplets \citep{Dupont2012}.

It is important to note that although these questions are crucial to study and predict the climate on Earth, most of these questions represent higher order uncertainties when dealing with exoplanets and paleo-climate problems. The cloud model described in this paper focuses on the first order climatic effects of clouds, and it represents the first radiative-convective, chemical, and cloud microphysical model developed for exoplanets in the habitable zone (HZ).


\subsection{The microphysical cloud model: convective water clouds}
\label{sec:liquid}
Microphysical cloud models were developed for Earth based on observational and laboratory data is available that constrain theoretical models \citep[see e.g., ][]{Rogers1989, Pruppacher1997, Khain2000, Seinfeld2006}. Two issues are important when developing a cloud model for exoplanets. 1) Calculating the vertical motion, condensation and coalescence of a droplet population self-consistently is computationally time consuming even in the 1D case. 2) Detailed observational data is not yet available for super-Earth and Earth-like exoplanets to constrain the initial conditions.

We therefore limit our microphysical model to resolve some basic properties of clouds without attempting to simulate cloud microphysics fully. These basic properties are the following: 
\begin{itemize}
\item The cloud base is located at the height where the local water mixing ratio equals the saturation mixing ratio depending on pressure and temperature.
\item As the convective parcel ascends, it cools and more water vapor condenses out. Therefore the size of droplets increases with height \citep{Warner1969, Rogers1989}.
\item The cloud top is located at the height where the relative velocity between the droplets is large enough to initiate coalescence. 
\end{itemize}
When coalescence becomes efficient, rain drops rapidly form and fall towards the surface. We do not directly simulate precipitation, only the formation of droplets because rain drops are removed from the atmosphere in a matter of a few minutes owing to their large terminal velocities \citep{Rossow1978}. Therefore the radiative effect of rain drops is negligible as they only spend a short amount of time in the atmosphere compared to the cloud droplets, which remain aloft.

The falling rain drops however influence the number density of cloud droplets as rain drops sweep up the cloud droplets as they fall through the cloud \citep{Pinsky1999}. Thus the number density of cloud droplets is reduced. This effect is considered in our model by the use of the precipitation efficiency parameter.

We consider two vertical columns. In one column, the air moves only upwards and clouds form by condensation. The other column undergoes downward motion, therefore it is expected to be cloud free.

The initial conditions for the cloudy column are the pressure-temperature structure and atmospheric composition in the convective zone ($T_i$ and $P_i$, where $i$ is the index of layers, $i=0$ being on the surface), relative humidity ($\chi_{\mathrm{surf}}$ or $\chi_0$) and the number density of aerosols ($n_{\mathrm{surf}}$) on the surface. From these quantities, we calculate the height of the cloud deck and top, the radius and number density of droplets as a function of height. We assume single-sized droplets at a given height, thus we do not follow the evolution of a droplet size distribution. This simplification is justified because the effective optical properties of a droplet size distribution is well represented by a single-sized droplet having the equivalent radius of the distribution \citep{Hu1993}.  

An air parcel on the surface becomes warmer than its environment, therefore it starts to rise and adiabaticly expands. If we assume that the air parcel does not exchange material with the surrounding air, the mixing ratio of water vapor will be constant as a function of height \citep[see however][on the importance of entrainment]{Houze1993}. The number density of aerosols on the other hand will decrease as the parcel expands:
\begin{equation}
n_i=n_{\mathrm{surf}} \frac{P_i}{P_{\mathrm{surf}}},
\label{eq:numdens}
\end{equation}
where $n_i$ is the number density of aerosols at pressure $P_i$ above the surface, $P_{\mathrm{surf}}$ is the pressure at the surface. The pressure scale height of the atmosphere is 10 km, therefore the number density of droplets only slightly decrease at the typical height of low level liquid water clouds, but the modification is significant for high cirrus clouds which form above 10 km.

When the mixing ratio of water in the parcel becomes larger than the saturation mixing ratio at the given temperature \citep[as described by][]{Rogers1989}, the excess water vapor condenses onto the surface of cloud condensation nuclei (CCN). The first layer where this happens marks the deck of the cloud layer. Above the cloud deck, the mixing ratio of water is equal to the saturation mixing ratio in that layer. The excess water vapor density in layer $i$ ($\rho_{ex,i}$) is the difference between the water vapor density in layer $i-1$ and the saturation vapor density in layer $i$. The vapor density is distributed uniformly amongst the droplets in the grid cell, and the mass of the water droplets in layer $i$ is
\begin{equation}
m_i = m_{i-1} + \rho_{ex,i}/n_i.
\end{equation}
Initially (below the cloud deck), the droplet mass is zero. The radius of the droplets as a function of height can be calculated and it increases with height in agreement with observations.

The Reynolds number of the droplets is
\begin{equation}
Re_i = \frac{2 r_i \rho v_{t,i}}{\mu_i},
\end{equation}
where $r_i$ is the droplet radius, $\rho$ is the density of water (1 g/cm$^{-3}$), $v_{t,i}$ is the terminal (or fall) velocity of the droplet \citep{Rossow1978}, and $\mu_i$ is the dynamic viscosity of the gas \citep{Rogers1989}. The Reynolds number describes the aerodynamical flow around the droplet. If it is small, the flow is laminar, high Reynolds number indicates a turbulent flow around the droplet. Once the flow becomes turbulent, the relative velocity between two droplets is increased and thus coalescence becomes efficient \citep{Rossow1978}. The critical Reynolds number above which coalescence starts is chosen to be $200$. Therefore, the cloud top is located at the height where the Reynolds number of the droplet becomes larger than 200. There are no droplets present above this height, as they would efficiently increase in radius due to collisions, and precipitate. The value of 200 was chosen because the droplet sizes at the cloud top is similar to the representative droplet sizes observed on Earth. We discuss the effects of different critical Reynolds numbers in Sect. \ref{sec:crit}.

Falling rain drops sweep up cloud droplets during their descent. An average cloud droplet has a size of $r_d = 10$ microns, an average rain drop is $r_D = 1$ mm. The mass ratio of a cloud droplet and a rain drop, assuming spherical particles, is $r_d^3 / r_D^3 =10^{-6}$, thus 10$^6$ cloud droplets must merge to produce a single rain drop. The number density of cloud droplets is 100 cm$^{-3}$, and the number density of rain drops is 10$^{-4}$ \citep{Rogers1989}. The ratio of number densities is also 10$^{-6}$, thus the precipitation efficiency is of the order of unity. As we do not follow the coalescence of a droplet distribution, we introduce a parameter called precipitation efficiency, $e_p$, which can vary between zero and unity. The number density of cloud droplets is reduced by a factor of $(1- e_p)$ due to precipitating rain drops.

The free parameters that determine this cloud layer are the surface relative humidity and CCN concentration, liquid water cloud fraction, and precipitation efficiency.

This convective cloud layer is self-consistently calculated in our 1D model and it is physically motivated. Such a cloud model provides a useful first order approximation on exoplanetary cloud properties, where the observational data in not abundant. However, our simulations show that for Earth, where detailed observations of cloud properties exist, the resulting convective cloud layer for humidity and CCNs cannot reproduce both the average surface temperature and global albedo of Earth. We find that a precipitation efficiency of 0.8 is necessary to obtain 289 K surface temperature, but the planetary albedo is only 0.21. A precipitation efficiency of 0.2 is needed to obtain a planetary albedo of 0.3, but the surface temperature is 275 K then. The reason why the surface temperature and albedo are not reproduced with one cloud layer is because the cooling effect of the low level water cloud dominates. If the single cloud layer were to form higher in the atmosphere, both the albedo and surface temperature could be reproduced. However, the only handle we have on the height of the convective cloud layer is through the surface relative humidity. Thus, in order to raise the height of the convective layer, we need to reduce the surface relative humidity and adopt a value which is different from what is observed on Earth.

In order to reproduce both the globally averaged surface albedo, temperature and the energy budget of Earth, we introduce a second cloud layer: cirrus clouds which provide additional warming effect.

\subsection{Cirrus clouds on Earth}
\label{sec:ice}
The cloud deck of cirrus clouds cannot be determined self-consistently in a 1D model because it is a result of horizontal motions (see Sect. \ref{sec:basic}). Therefore we use a parametrization to determine the height of the cloud layer. Cirrus clouds on Earth form typically at temperatures of 230 K, which corresponds to a height of 12 km \citep{Liou1986, Dowling1990}. There are two possible ways to proceed. One can either scale the cirrus height relative to the height of the troposphere. The other possibility is to place the cloud deck where the temperature is 230 K. Clouds do not form above the troposphere, thus the advantage of the former method is that it expresses the cloud height as a function of its natural maximum value. The advantage of the latter method is that a fixed cloud temperature corresponds to a fixed super saturation relative to water ice, and it also directly informs us about the greenhouse effect of these clouds. Neither of the two methods have an overwhelming advantage over the other. However, we found in our simulations that linking the cloud height with the tropopause height sometimes results in non-convergence and continuous surface cooling. Therefore, for pure numerical reasons, we use the latter method to determine the height of the cloud deck.

The size of the particles as a function of height is determined in the next step. We assume that the air parcel is saturated relative to liquid water, thus super-saturated relative to ice. This assumption is commonly used in cirrus cloud models \citep{Starr1985, Starr1985a, Liou2000}. The difference condenses onto ice condensation nuclei (ICN) and the particle sizes as a function of height can be determined. The number density of ICN is calculated from the same surface aerosol number density as for liquid water clouds for simplicity.

The height of the cloud top is reached where the particle Reynolds number is larger than 200, as for liquid water droplets. The falling particles sweep up other particles as they settle thus reducing the particle number density. Again, for simplicity we assume the same precipitation efficiency as for liquid water clouds.

As a result of our assumptions, we only need one additional free parameter (the ice cloud fraction) to describe the second cloud layer. \\

Once the particle properties in the two cloud layers are determined, we calculate the visible and IR fluxes. We use flux averaging to determine the equilibrium pressure-temperature structure for a cloudy atmosphere \citep{Kitzmann2010, Goldblatt2011}. The initial or previously determined T-P profile is used to calculate the radiative fluxes. The T-P profile and mixing ratios are kept fixed and the incoming and outgoing fluxes are calculated for four cases: cloud-free T-P profile ($F_{cf}$), T-P profile with liquid water clouds only ($F_{lc}$), with ice clouds only ($F_{ic}$), and with both clouds overlapping ($F_o$). These four fluxes are averaged using the cloud fractions as weight and assuming a random overlap between the two cloud layers:
\begin{eqnarray}
F = F_{cf}\times (1 - f_l - f_i + f_l f_i) + F_{lc} \times (f_l - f_l f_i) +\nonumber \\ F_{ic}\times (f_i - f_l f_i) + F_o\times (f_l f_i),
\end{eqnarray}
where $f_l$ is the cloud fraction of the liquid water cloud, and $f_i$ is the cloud fraction of the ice cloud. The average flux ($F$) is used in Eq. \ref{eq:dTdt} to calculate the new, updated T-P profile.

\begin{figure}
\centering
  \includegraphics[width=0.49\textwidth]{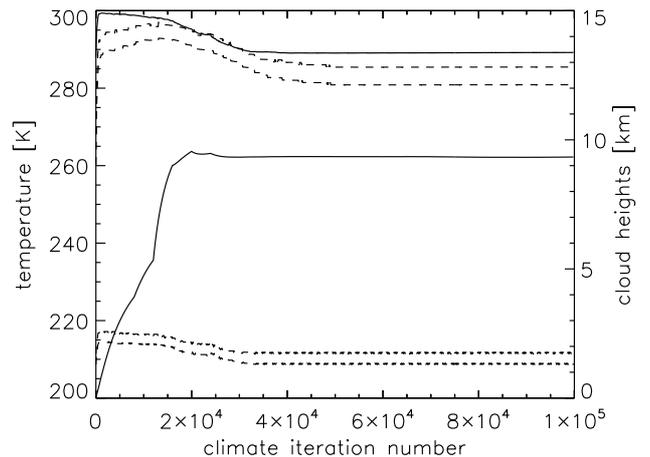}
  \caption{Convergence of the surface temperature, stratosphere temperature and cloud heights. The x axis shows the number of iterations performed by the climate code. The y axis (left side) is the temperature. The upper solid line shows the surface temperature, the lower solid line shows the stratosphere temperature at $10^{-3}$ bars, where the stratosphere temperature is the highest.  The y axis (right side) is height in km, where the lower pair of dotted lines show the height of the liquid water cloud. The upper pair of dotted lines show the ice cloud heights. Convergence in all of these quantities is reached after $5\times10^{4}$ climate iterations.}
  \label{fig:conv}
\end{figure}

\section{Model set up and tests}
\label{sec:tests}
\subsection{Initial conditions}
\label{sec:inicond}
Unless otherwise stated, we use the following initial conditions for all simulations performed in this paper. We adopt the physical and orbital parameters of the Earth-Sun system and initiate the simulations with the standard Earth atmosphere composition: $f_{Ar}=10^{-2}$, $f_{CH_4}=1.6\times 10^{-6}$, $f_{CO_2}=3.55\times 10^{-4}$, $f_{O_2}=0.21$, where $f$ is the volume mixing ratio of argon, methane, carbon-dioxide, and molecular oxygen respectively. The rest of the atmosphere consists of molecular nitrogen ($f_{N_2}=0.77$). Other important tracer species are water vapor, and ozone whose initial mixing ratio is set to $f_{H_2O}=10^{-5}$ in the stratosphere, and $f_{O_3}=10^{-6}$. The relative humidity is a free parameter in the code. However, we set it to 0.77 on the surface (global average Earth value) for the convergence tests described in Sect. \ref{sec:conv_test}. Most other chemical species are initiated with $f_{\mathrm{other}}=10^{-20}$.

The initial temperature and pressure profile is calculated as follows. The surface pressure is set to 1 bar, the pressure on the top of the atmosphere is $5 \times 10^{-5}$ bar representing about 70 km in current Earth's atmosphere. The pressure grid is initialized such that it is uniformly spaced in log-pressure coordinates (but see Par. 3 in Sect. \ref{sec:conv_test}). The temperature on the surface is set to 289 K, the temperature in the stratosphere is initially 200 K. The height of the troposphere is set to 10 km (but evolves during the simulation) and the temperature in the troposphere drops linearly to reach 200 K at the top of the troposphere. Once the pressure and temperature is known, the height of the layers can be calculated assuming hydrostatic equilibrium and using the ideal gas equation of state.

As discussed in Sect. \ref{sec:changes}, the surface albedo of the planet was initially set to $\alpha_{\mathrm{surf}}=0.22$ in the cloud-free model to mimic the effects of clouds. However, the clear sky albedo of Earth is 0.15 \citep[Chapter 5 of][based on ERBE measurements]{Pierrehumbert2011}. The clear sky albedo is the combined albedo of the gaseous atmosphere (reflecting light via Rayleigh scattering) and the reflectance of the surface, no clouds are considered. The clear sky albedo in the model can be measured as the ratio of the outgoing and incoming visible/near-IR flux on the top of the atmosphere ($F_{\mathrm{\uparrow top}}/F_{\mathrm{\downarrow top}}$). We set the surface albedo such that $F_{\mathrm{\uparrow top}}/F_{\mathrm{\downarrow top}}=\alpha_{\mathrm{clear}}=0.15$. If we follow this procedure, we find that the surface albedo has to be set to $\alpha_{\mathrm{surf}} = 0.13$ \citep[in agreement with][]{Trenberth2009}. This means that 2\% of the incoming radiation is reflected off the atmosphere by atoms and molecules in the clear sky case. The surface temperature is then between 292 and 295 K depending on the number of vertical layers considered (see Sect. \ref{sec:conv_test}). As this value is larger than the global average of 289 K, it already suggests that cooling effect of water clouds dominates over the warming effect of cirrus clouds on Earth. The zenith angle is set to $60^\circ$ in these simulations. 

The time step in the climate model is adaptive and it is regulated by the maximum temperature difference in all layers ($\Delta T$). It is increased by a factor of 1.1, if $\Delta T$ is smaller than $5 \times 10^{-2}$ K, the time step is divided by two, if $\Delta T$ is greater than $10^{-1}$ K. These values are an order of magnitude smaller than in the clear sky model due to the finer vertical resolution in our model with clouds.

Clear sky calculations call the climate code 200 times, the photochemistry code 400 times, and the whole procedure is repeated 2-3 times until convergence is reached \citep{Segura2010}. The cloudy sky model calls the climate code 4000 times, the photochemistry 800 times, and repeats the procedure 20 times. The significant increase in iteration number is necessary because of the small time steps and different initial conditions. We illustrate the convergence of surface temperature, stratosphere temperature, and cloud heights on Fig. \ref{fig:conv}. The cloud model parameters of the simulation are described in the third convergence test of Sect. \ref{sec:conv_test}. We see that both the surface temperature and cloud heights reach convergence after $5\times10^{4}$ climate iterations. However, in some cases where the T-P profile differs significantly from the initial T-P profile, convergence is reached later, thus iterating $10^5$ times is justified. 

\begin{figure*}
\centering
  \includegraphics[width=0.47\textwidth]{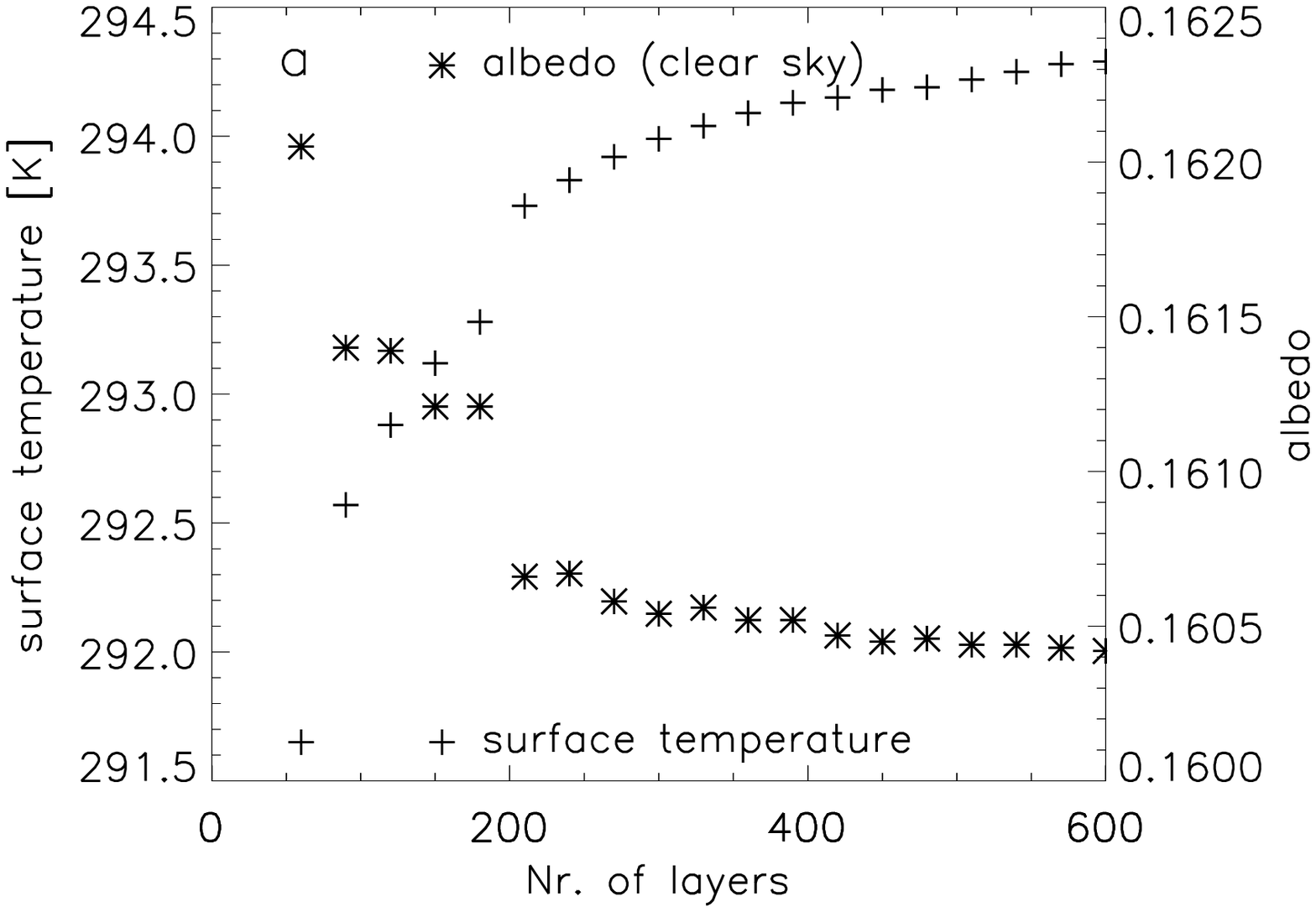}
  \includegraphics[width=0.47\textwidth]{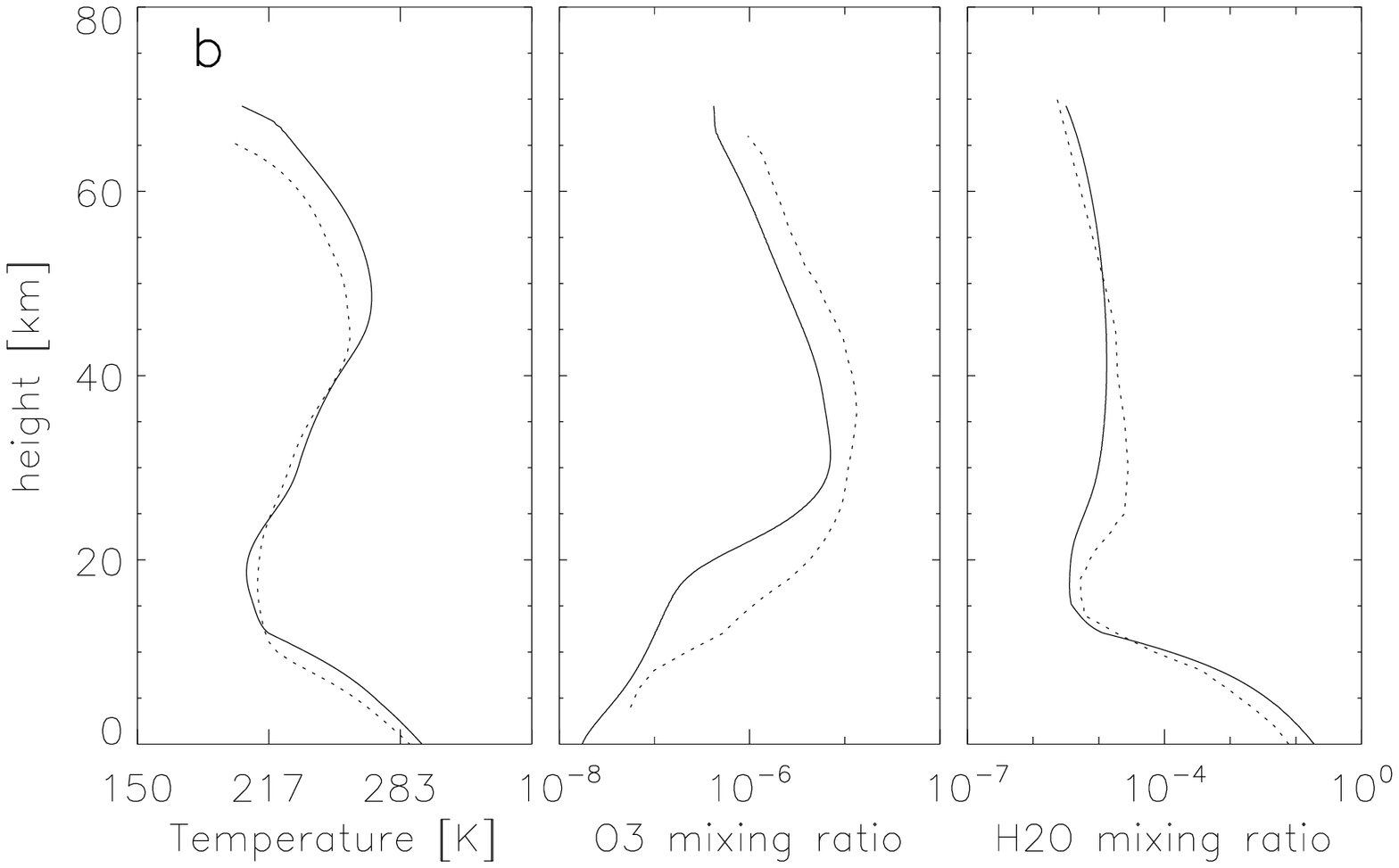}
  \caption{\emph{Left side:} convergence test for the clear-sky case. The solid line shows the surface temperature as a function of the number of layers used in the model. The dotted line shows the surface albedo. \emph{Right side:} the temperature, ozone and water mixing ratios as a function of height for $n_z=600$. The dotted lines show the \cite{USatmo1976} profiles.}
  \label{fig:test_clear}
\end{figure*}

\begin{figure*}
\centering
  \includegraphics[width=0.47\textwidth]{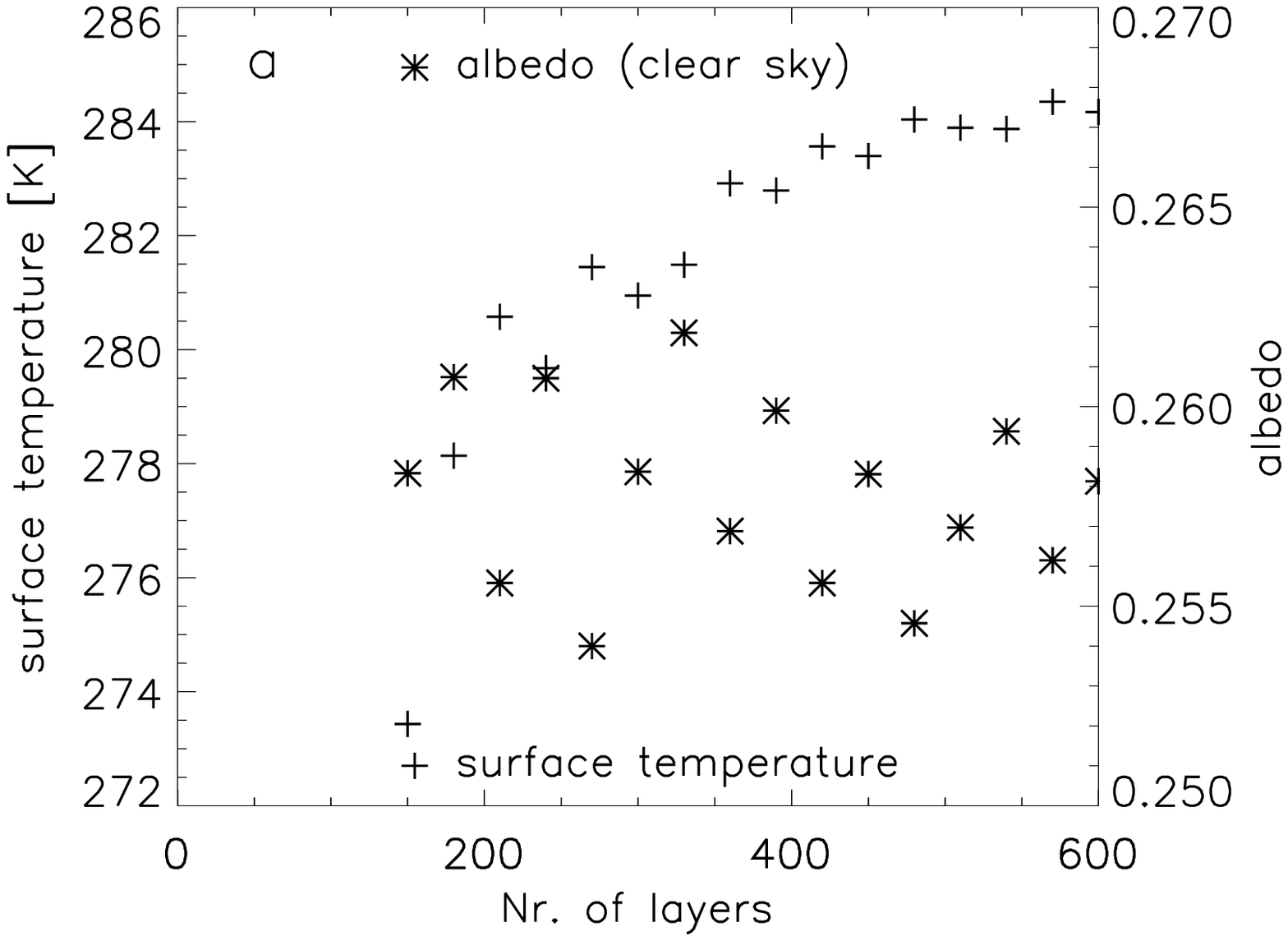}
  \includegraphics[width=0.47\textwidth]{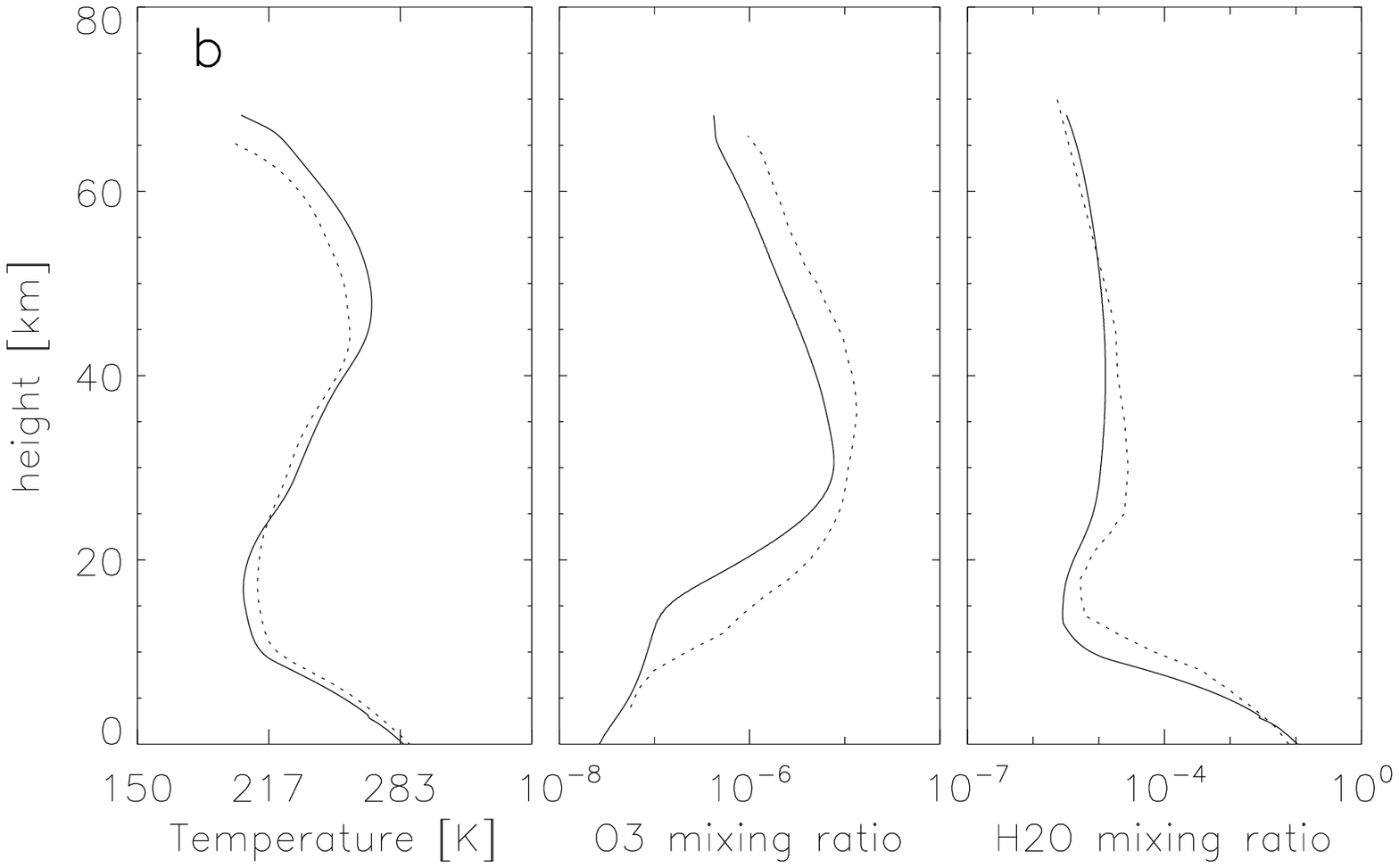}
  \caption{\emph{Left side:} the surface temperature and albedo as a function of layer number used for the parameterized cloud convergence test. We assume a 40\% cloud coverage, the cloud deck and top are located at $P=0.85$ and 0.7 bars, respectively, the droplets have a radius of 11 microns, and the total water path is 40 g/m$^2$ \citep{Goldblatt2011}. \emph{Right side:} the temperature, water and ozone mixing ratio profiles as a function of height for $n_z=600$. The dotted lines show the \cite{USatmo1976} profiles.}
  \label{fig:test_simple_c}
\end{figure*}

\begin{figure*}
\centering
  \includegraphics[width=0.47\textwidth]{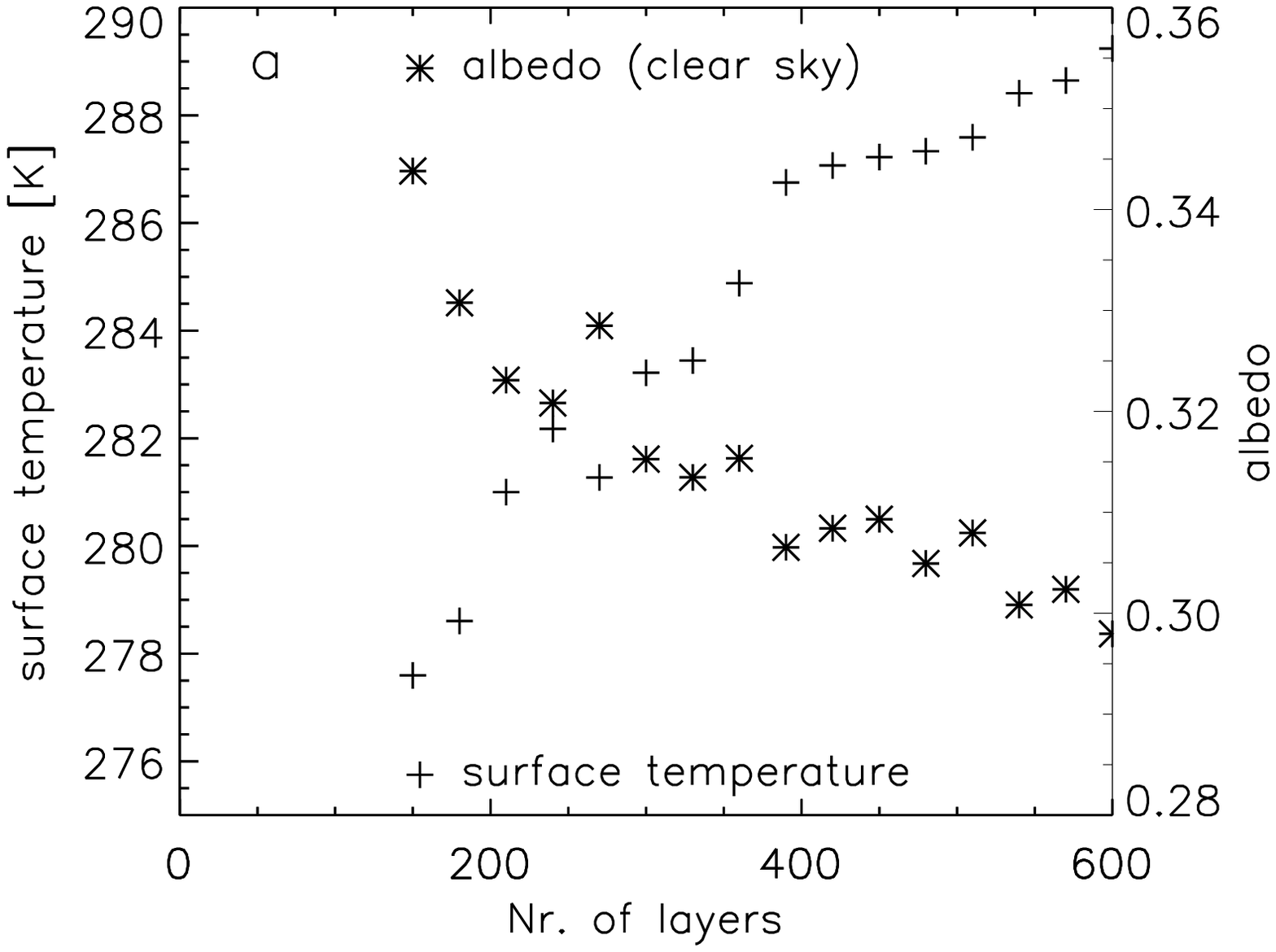}
  \includegraphics[width=0.47\textwidth]{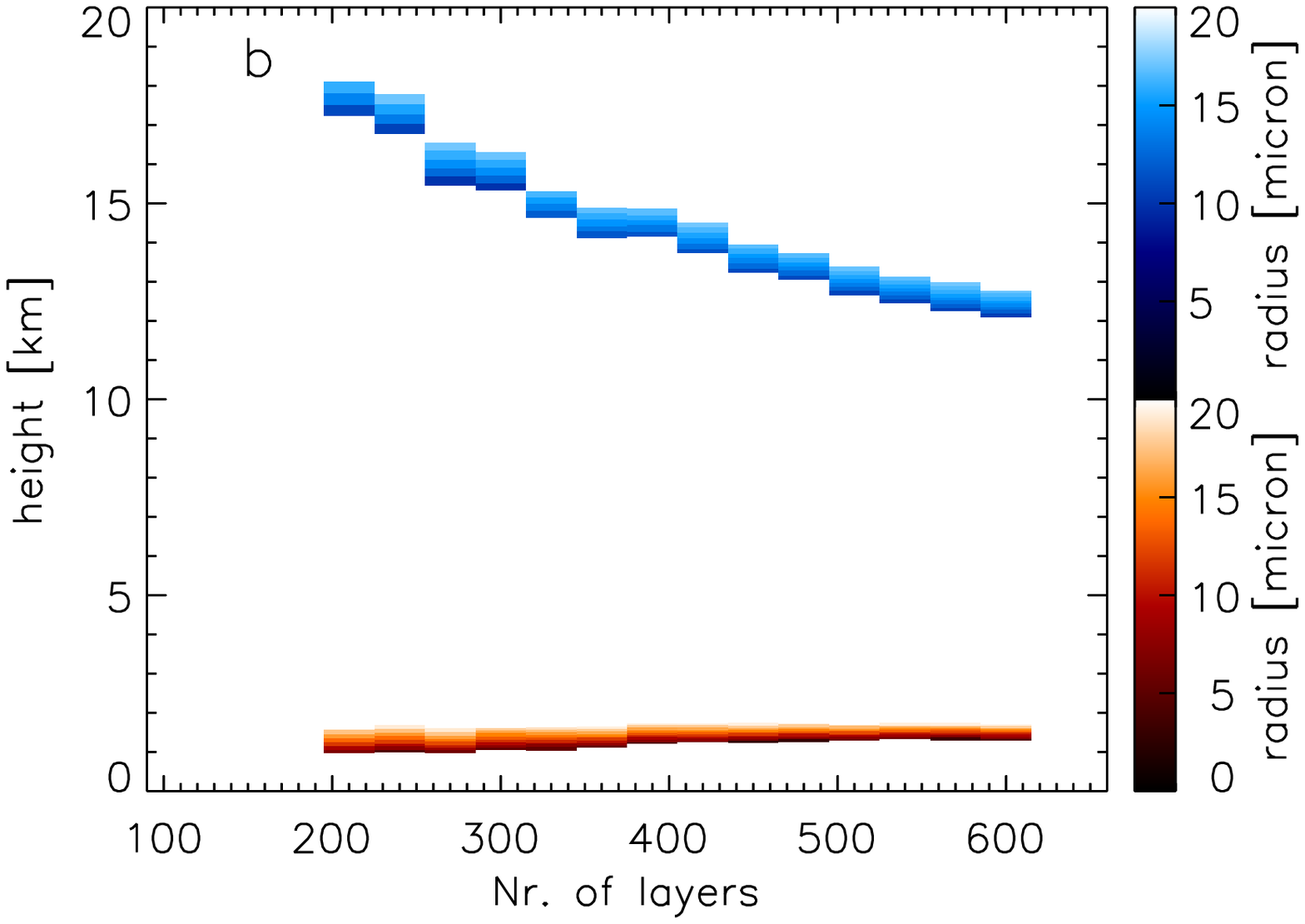}
  \caption{\emph{Left side:} the surface temperature and albedo as a function of layer number used for the convergence test of the microphysical cloud model. We assume a surface relative humidity of 77\%, an aerosol number density of 100 cm$^{-3}$, a liquid water cloud fraction of 40\%, cirrus cloud fraction of 25$\%$, and a precipitation efficiency of 0.8. \emph{Right side:} the properties of the cloud layers. The y axis shows the height of the layer, the color represents the radius of the droplets in the layers. The red contours show the liquid water cloud properties, the blue contours correspond to the ice cloud.}
  \label{fig:test_micro_c}
\end{figure*}

\subsection{Convergence tests}
\label{sec:conv_test}
\paragraph{Clear-sky test} Using the above mentioned initial conditions, we performed clear-sky simulations (no clouds) with increasing number of grid cells to verify the convergence of the code. The number of layers varies between $n_z=60$ to 600 with a step size of 30. Figure \ref{fig:test_clear}a shows the surface temperature and clear sky albedo as a function of layer numbers. As expected, both the surface temperature and the clear-sky albedo converges as the number of vertical layers increases. The surface temperature difference between $n_z=570$ and 600 runs is 0.01 K, the surface albedos are basically equal (a difference occurs only at the fifth decimal point). Therefore we conclude that convergence is reached for $n_z=600$.

Figure \ref{fig:test_clear}b shows the temperature, ozone and water mixing ratios as a function of height after convergence was reached for $n_z=600$. The dotted lines show the \cite{USatmo1976} profiles. The agreement between the standard atmosphere profiles and the derived profiles are not perfect in Fig. \ref{fig:test_clear} for two reasons. 1) The standard atmosphere profiles are based on globally averaged measurements, therefore these profiles might not represent an atmosphere that is in radiative and chemical equilibrium. 2) The correct Earth surface temperature is influenced by clouds, which is not accounted for in this model atmosphere. 

\paragraph{Parameterized cloud model} We adopt a homogenous cloud model similar to the one used by \cite{Goldblatt2011} for their low level water clouds. We assume a 40\% cloud coverage, cloud deck at 0.85 bar, cloud top at 0.7 bar, mono-disperse droplets meaning that all droplets have the same radius, 11 micron, and column density of cloud water (or water path) 40 g/m$^2$. The number of layers are varied between 150 and 600 with a step of 30. We found that using less than 150 layers introduce larger uncertainties in the surface temperature and albedo because there are only one or two cloudy layers, which can potentially extend to a much wider or narrower pressure range than 0.85 and 0.7 bar. The number of cloudy layers are three and twelve for 150 and 600 vertical layers, respectively.

The surface temperature and albedo are shown on Fig. \ref{fig:test_simple_c}a, the temperature, ozone and water mixing ratios are shown on Fig. \ref{fig:test_simple_c}b. The surface temperature is between 273-285 K, the albedo of the planet is between 0.25-0.27 depending on the number of layers used. Convergence is observed for the surface temperature and for the planetary albedo. The cloud deck and top are not exactly at the prescribed pressures due to the vertical discretization of the pressure grid. As the exact pressure values of the cloud deck and top fluctuates, so does the surface temperature and albedo values as seen on Fig. \ref{fig:test_simple_c}.

\paragraph{The microphysical cloud model} In this convergence test, we use the microphysical cloud model as described in Sects. \ref{sec:liquid} and \ref{sec:ice}. The free parameters of the microphysical cloud model are the number density of aerosols ($n_{\mathrm{surf}}$), the cloud fractions ($f_l$, and $f_i$), the relative humidity on the surface ($\chi_{\mathrm{surf}}$), and the precipitation efficiency ($e_p$). Here we use $n_{\mathrm{surf}}=100$ cm$^{-3}$ \citep{Miles2000}, $\chi_{\mathrm{surf}}=0.77$, which both are average measured Earth quantities. Finally, $e_p=0.8$, $f_l=0.4$, and $f_i = 0.25$ are used to obtain surface temperatures equal to the measured average Earth value of 289 K. 

In a pressure grid which is uniformly spaced in log pressure (see Sect. \ref{sec:inicond}), there are only two cloudy layers even for 600 grid cells, thus we found no real convergence. One could increase the number of layers above 600, but that renders the computations too expensive. We change the pressure grid instead. The ratio of grid spacing at the top to the grid spacing at the bottom of the atmosphere is set to 15, which means that the lower atmosphere (where clouds are located) is well resolved, the grid spacing at the top of the atmosphere is coarse. Under such conditions, there are on average 4 and 10 cloudy grid cell per cloud layer for 150 and 600 grid cells, respectively (see Fig. \ref{fig:test_micro_c}b). 

Figure \ref{fig:test_micro_c}a shows the surface temperature and albedo as a function of layer number. The surface temperature varies between 277 and 289 K, while the albedo varies between 0.345 and 0.3. The albedo decreases and surface temperature increases with greater layer numbers. Figure \ref{fig:test_micro_c}b shows the properties of the clouds in the simulations. The y axis shows the height of the cloudy layers, the center of the layers is marked with a `+' sign, and the colors correspond to the droplet radius. Red contours show the properties of the liquid water cloud, the blue contours show the ice cloud properties. In agreement with observation and theory, the droplet size is increasing with height \citep[e.g.,][]{Warner1969, Rogers1989}. There are no droplets with a radius above 20 microns at the adopted pressure and temperature conditions because of rain-out. 

We use 600 layers and a grid spacing ratio of 15 for the following simulations.

\subsection{CO$_2$ doubling}
We perform one simulation to measure the climate sensitivity due to CO$_2$ doubling. This is another way to validate our model. We use the standard cloud parameters for Earth (see Sect. \ref{sec:preceff} with precipitation efficiency 0.8) but with a doubled CO$_2$ concentration ($f_{CO_2}=7.1 \times 10^{-4}$). We find that the surface temperature is 292.5 K, and the planetary albedo is 0.29. The present Earth values are 289 K and 0.3. Both cloud layers form at slightly increased altitudes. The cloud decks are located at 1.6 km and 13.1 km. These values in our present Earth simulation are 1.3 km and 12.1 km.

The climate sensitivity of our model to CO$_2$ doubling is thus 3.5 K which compares favorably to the IPCC best estimate of 3 K with a likely range of 2 - 4.5 K \citep{IPCC2007}. Therefore we conclude that our model reliably reproduces the climate of Earth.

\section{Results}
\label{sec:res}
We perform five sets of simulations and in each of these sets, we keep the value of four parameters constant and vary only the fifth one. We could map the whole parameter space by simulating all possible combinations of the free parameters, but we feel that our understanding is better served by varying only one parameter at a time and exploring the effects individually.

\begin{figure*}
\centering
  \includegraphics[width=0.45\textwidth]{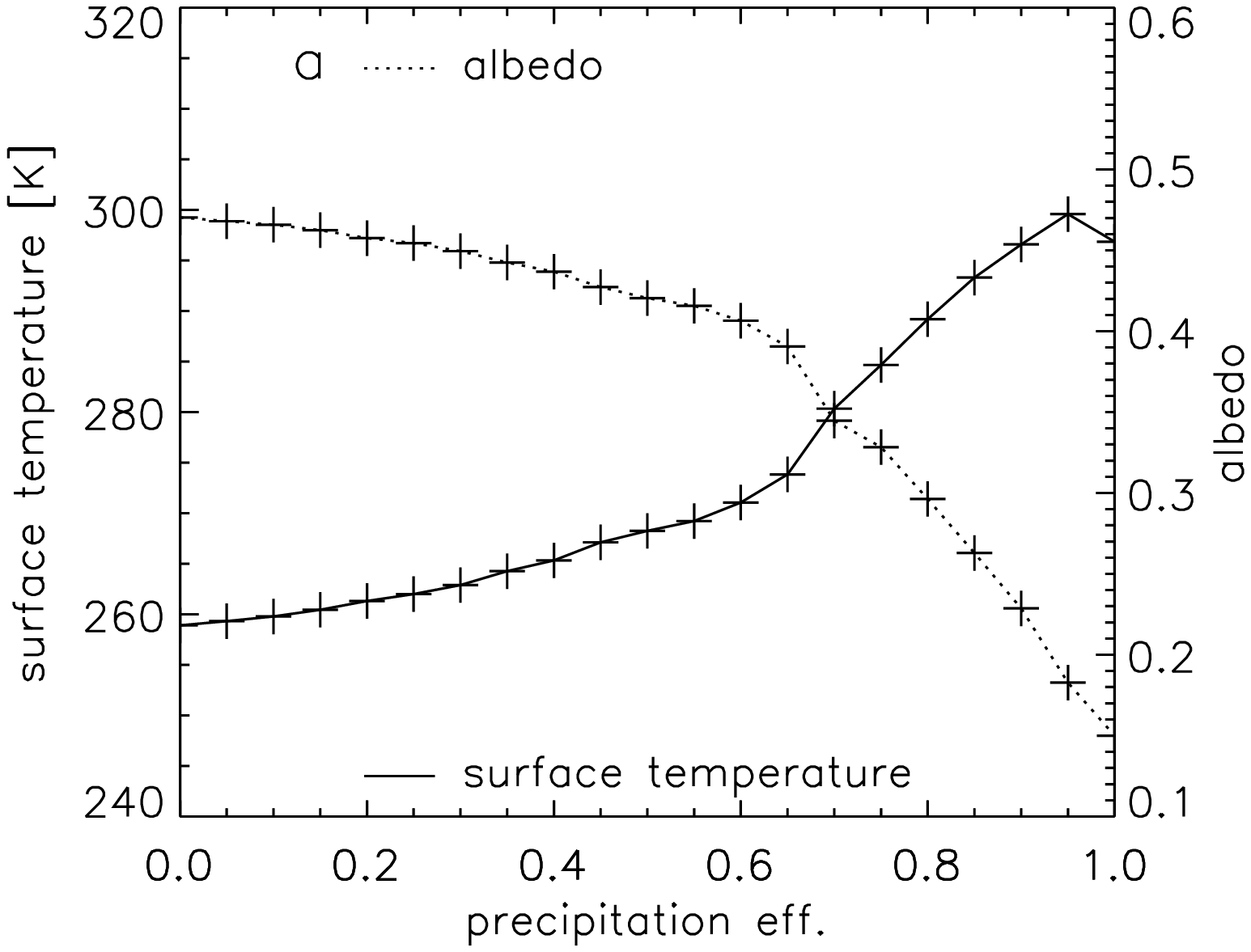}
  \includegraphics[width=0.45\textwidth]{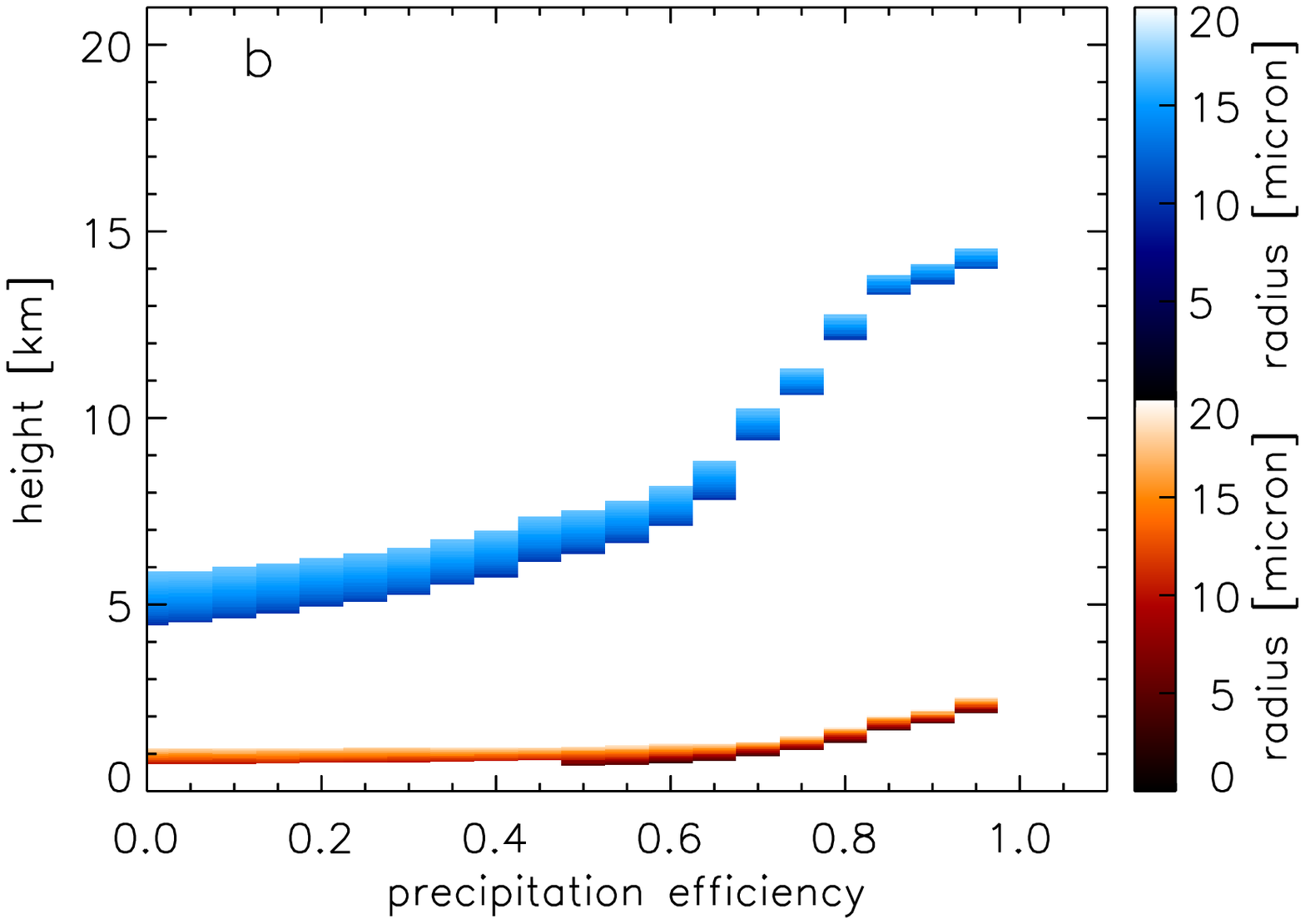}
  \caption{The precipitation efficiency is varied between 0 and 1 for $f_l=40\%$, $f_i=25\%$, aerosol number density of 100 cm$^{-3}$, and relative humidity of 77\%. \emph{Left side:} the albedo and surface temperature as a function of precipitation efficiency. \emph{Right side:} the height of the cloudy layers with the `+' signs. The color indicates the droplet sizes.}
  \label{fig:res_preceff}
\end{figure*}

\subsection{Precipitation efficiency}
\label{sec:preceff}
First, we set the number density of aerosols to 100 cm$^{-3}$, the relative humidity to 77\%, the liquid water cloud fraction to 0.4, and the ice cloud fraction to 0.25 in agreement with \cite{Goldblatt2011}, and vary the precipitation efficiency between 0 and 1 with a step of 0.05. This set of simulations is used to determine what precipitation efficiency is needed to reproduce the average surface temperature, albedo and GEB of Earth by adopting the measured values of the other four cloud parameters.

We show the albedo and the surface temperature as a function of precipitation efficiency on Fig. \ref{fig:res_preceff}a. A low precipitation efficiency produces geometrically and optically thick clouds and the albedo effect of clouds dominates over their greenhouse effect. As the precipitation efficiency is increased, the clouds become more optically thin, more of the incoming solar radiation reaches and warms the surface.
For a precipitation efficiency of unity, there are no clouds present and we obtain the clear sky albedo and surface temperature values. The precipitation efficiency value resulting in 289 K surface temperature is 0.8, and the corresponding albedo is 0.3 in agreement with observations \citep[Chapter 5 of][based on ERBE measurements]{Pierrehumbert2011}.

The cloud properties are illustrated on Fig. \ref{fig:res_preceff}b, which shows the height of the cloudy grids as a function of precipitation efficiency with `+' signs, and the colors represent the droplet radius in microns separately for the liquid and ice clouds. As the surface temperature increases, so does the height of the cloudy grid cells. The size of the droplets increases with height in accordance with observations and theory and their maximum size does not exceed 20 microns due to the onset of precipitation. 

The GEB describes two processes: 1) how the incoming solar flux is absorbed by atoms, reflected by clouds, how much solar flux reaches the surface, leaves the surface and finally the atmosphere; 2) the amount of thermal IR flux emitted by the surface, how much IR flux is back radiated to the surface, and leaves the atmosphere. The GEB of Earth is determined by satellite observations \citep{Zhang2004} and we compare the GEB of our simulation to the observed values in Tab. \ref{table:GEB} for a precipitation efficiency of 0.8. The maximum relative error of the observed and measured quantities is $6\%$, the typical relative error is $3\%$, thus our model reproduces Earth's GEB with a precipitation efficiency of 0.8. 

\begin{figure*}
\centering
  \includegraphics[width=0.45\textwidth]{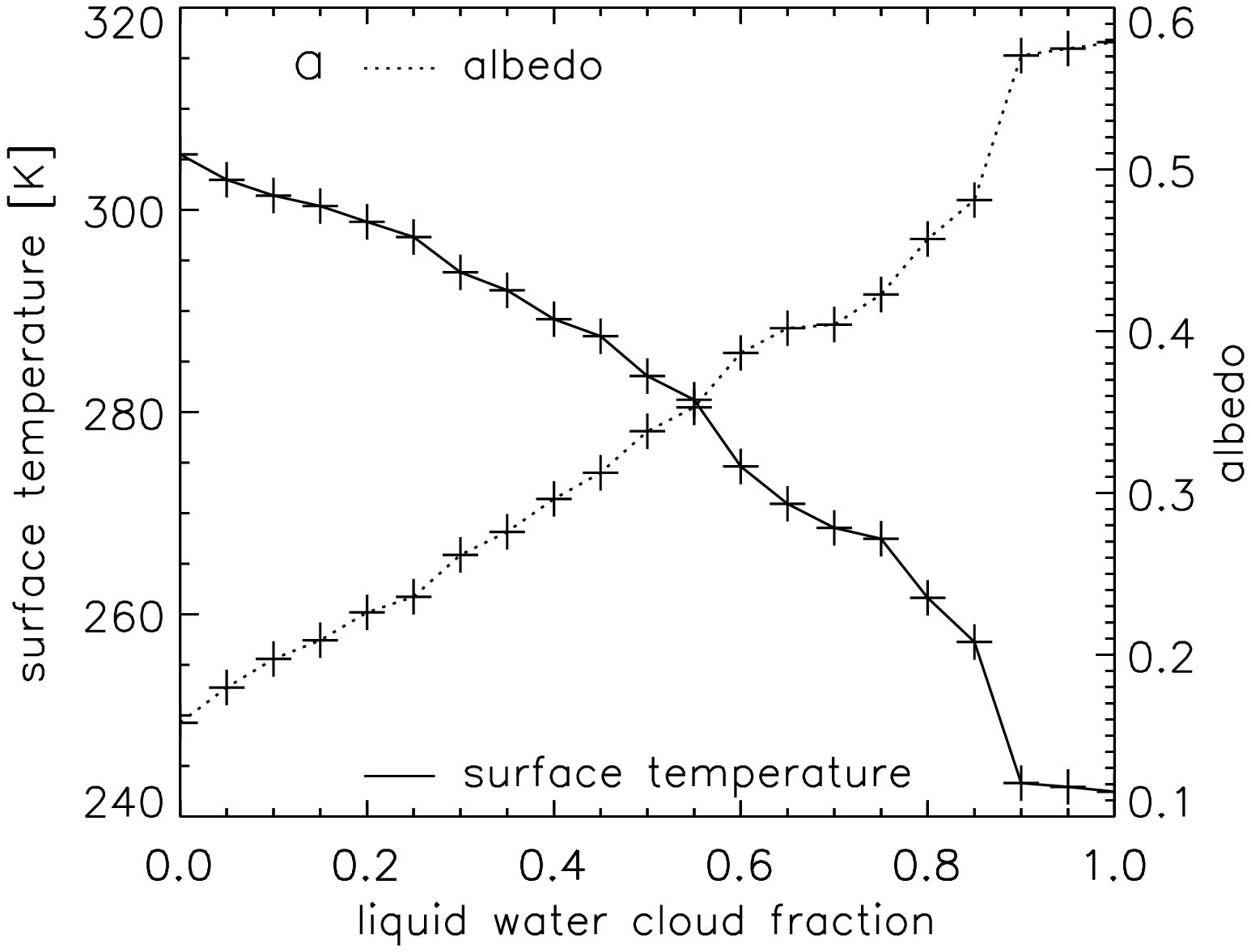}
  \includegraphics[width=0.45\textwidth]{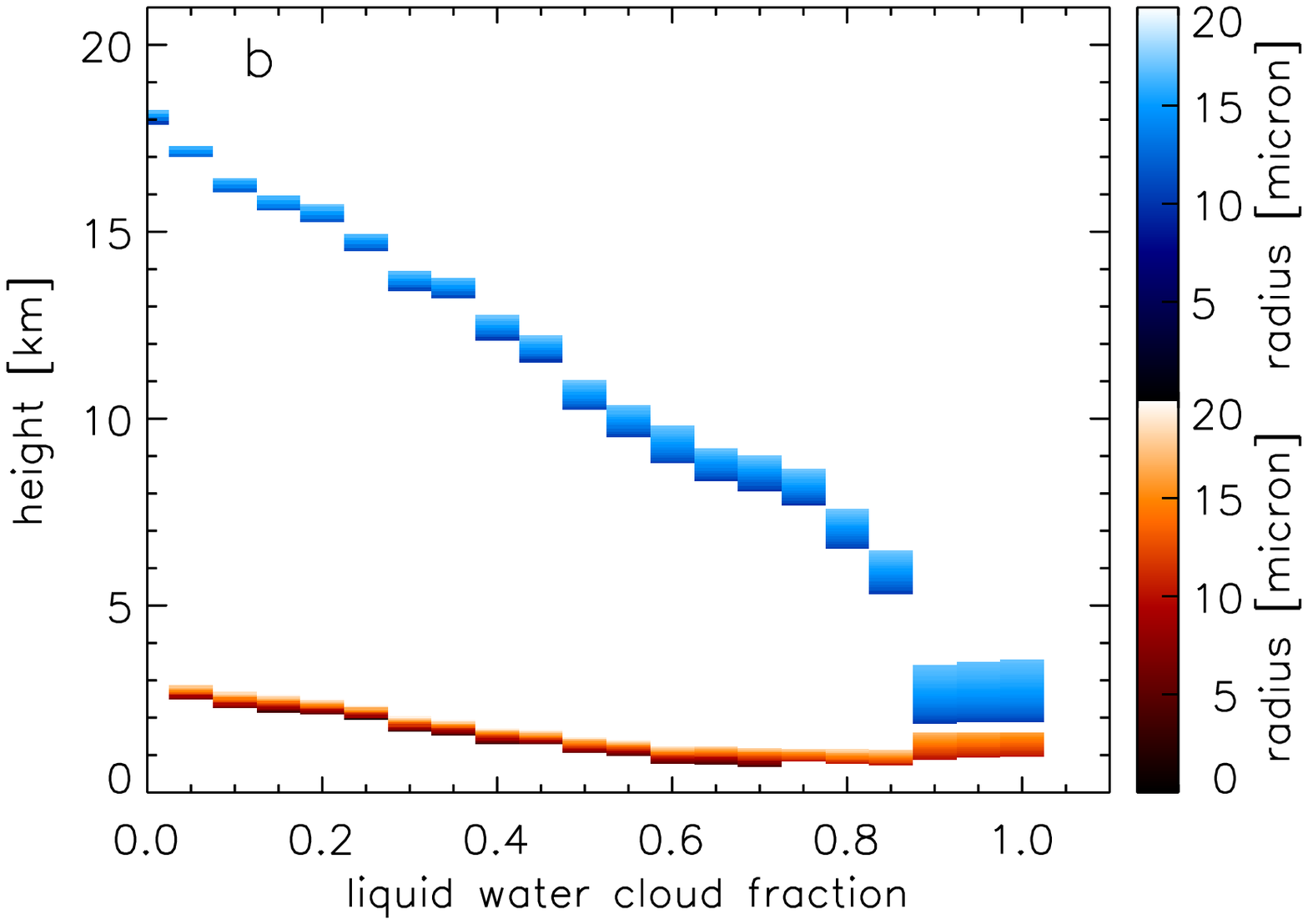}
  \caption{The liquid water cloud fraction is varied between 0 and 100\% for $f_i=25\%$, an aerosol number density of 100 cm$^{-3}$, relative humidity of 77\%, and precipitation efficiency of 0.8. \emph{Left side:} the albedo and the surface temperature as a function of cloud fraction. \emph{Right side:} the height of the cloudy layers and the size of the droplets.}
  \label{fig:res_cfl}
\end{figure*}

\begin{figure*}
\centering
  \includegraphics[width=0.45\textwidth]{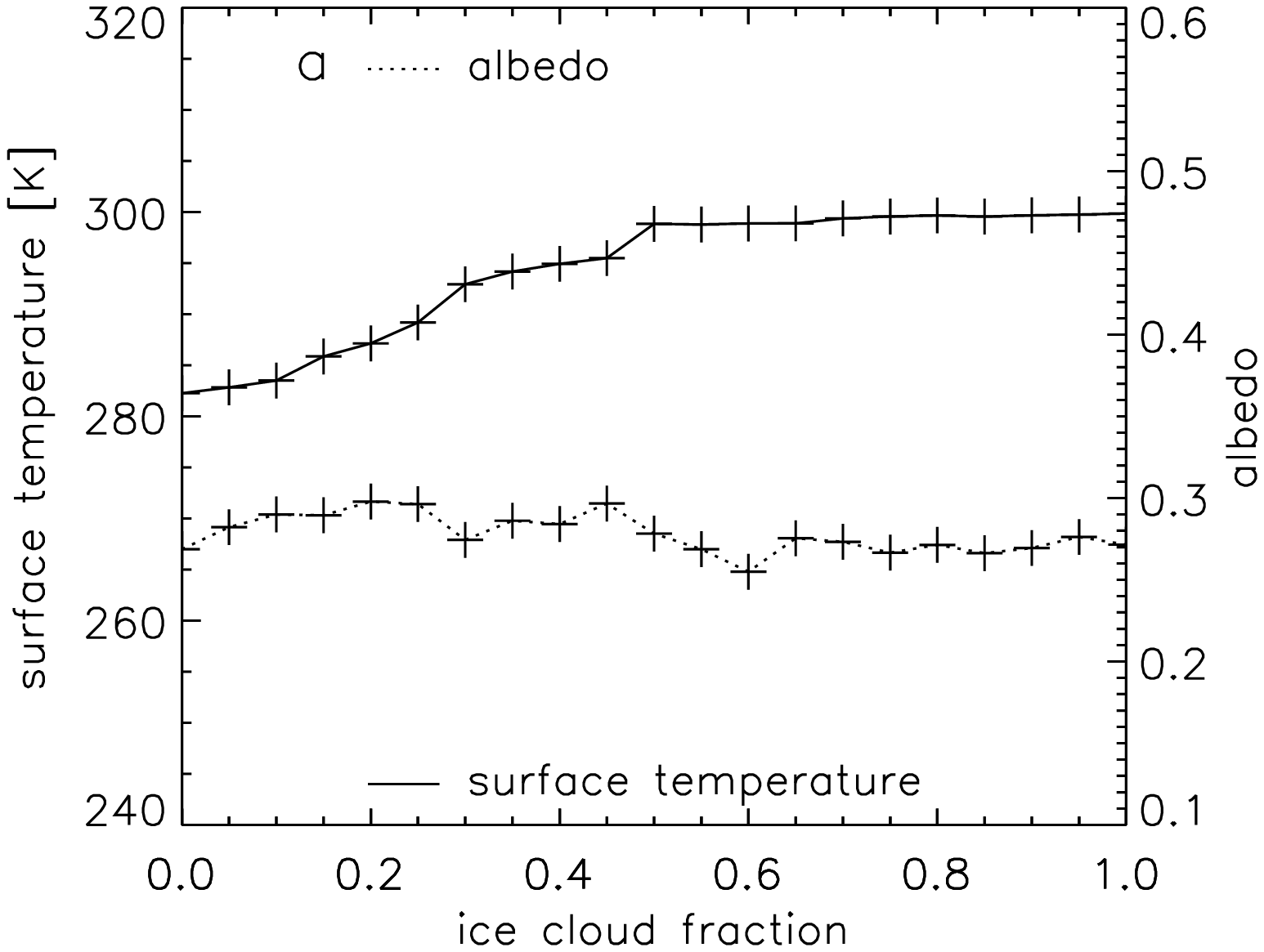}
  \includegraphics[width=0.45\textwidth]{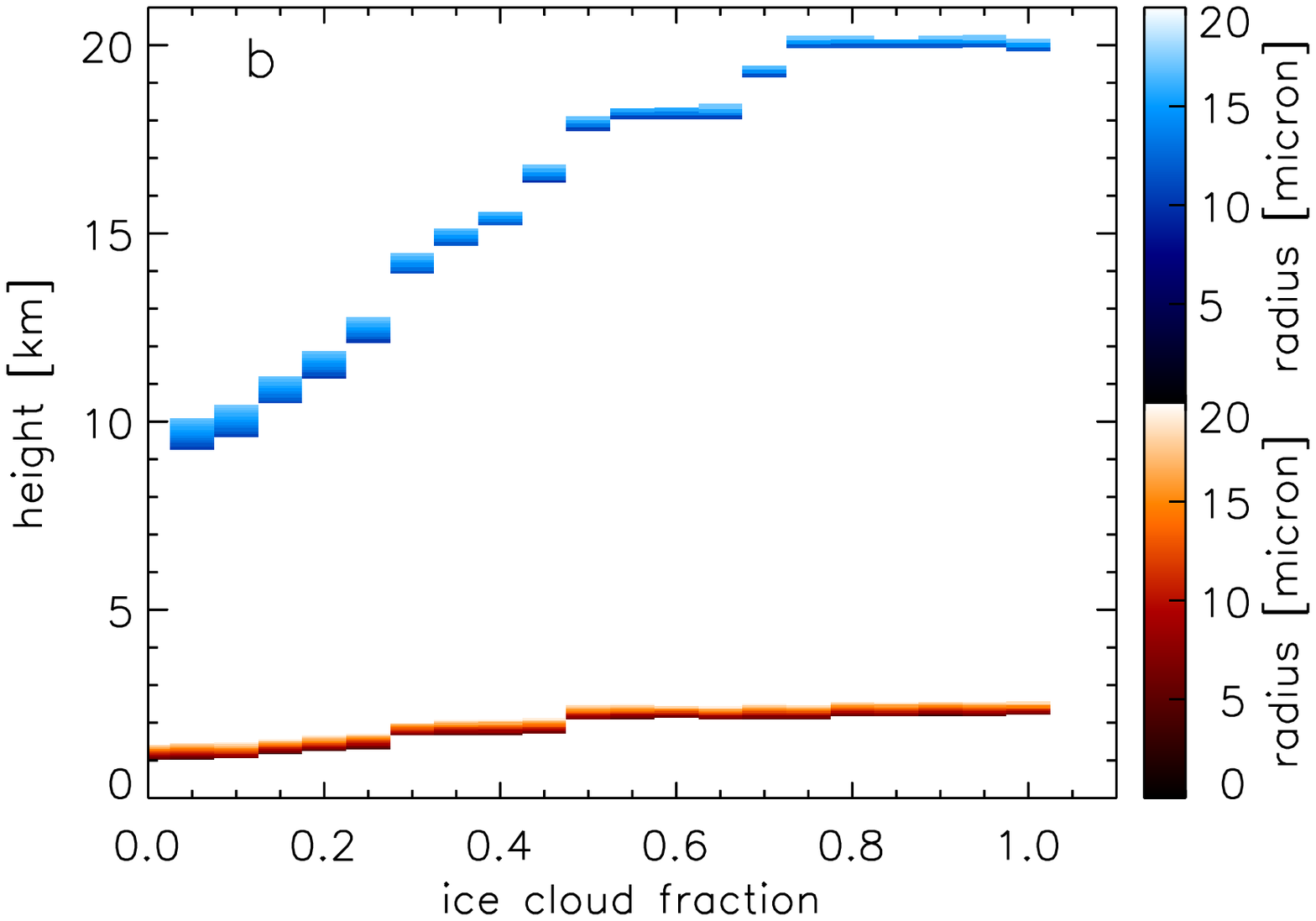}
  \caption{The water ice cloud fraction is varied between 0 and 100\% for $f_l=40\%$, an aerosol number density of 100 cm$^{-3}$, relative humidity of 77\%, and precipitation efficiency of 0.8. \emph{Left side:} the albedo and the surface temperature as a function of cloud fraction. \emph{Right side:} the height of the cloudy layers and the size of the droplets.}
  \label{fig:res_cfi}
\end{figure*}

\begin{table}
\caption{Global energy budget. The first column describes the flux type, the second column is the observed flux values from \cite{Zhang2004}, the third column is the flux values calculated in our model to reproduce Earth's GEB. The relative error between the measured and simulated quantities is a few percent at maximum, therefore we conclude that the agreement is good.}
\begin{minipage}[t]{\columnwidth}
\label{table:GEB}      
\centering
\begin{tabular}{l r r}        
\hline\hline                 
&Measured&Simulation\\
&[W/m$^2$]&[W/m$^2$]\\
\hline                        
Incoming solar flux					&341.8	&340.3\\
Absorbed by atoms					&70.9	&73.1\\
Reflected by atoms and clouds			&81.7	&76.6\\
Reached the surface					&189.2	&190.6\\
Leaves the surface					&24.0	&24.8\\
Leaves the atmosphere				&105.7	&101.4\\
\hline
Emitted IR flux from the surface		&395.6	&396.8\\
Back radiation to the surface			&344.7	&344.8\\
IR flux leaving the atmosphere			&233.3	&238.8\\
\hline
\end{tabular}
\end{minipage}
\end{table}

\begin{figure*}
\centering
  \includegraphics[width=0.45\textwidth]{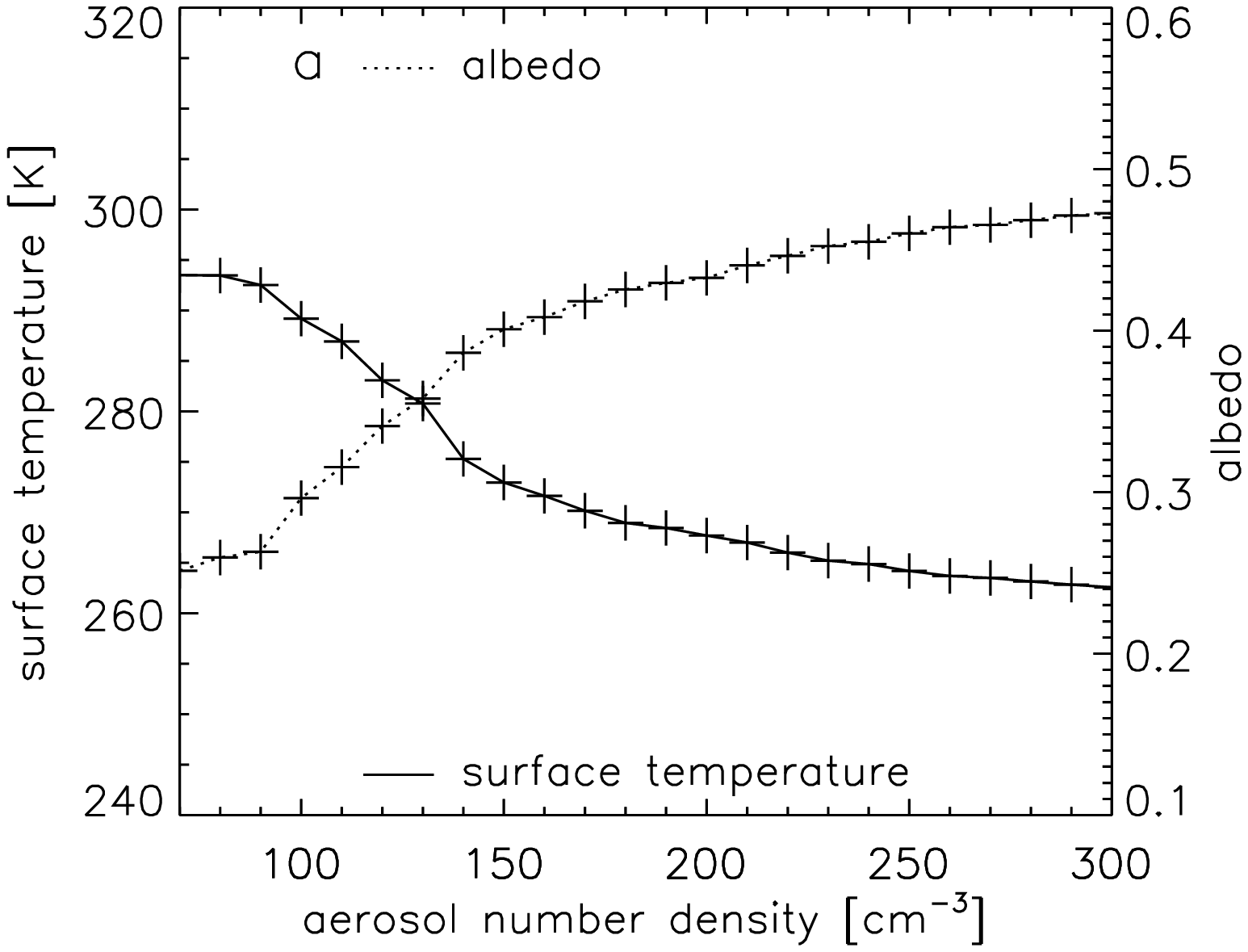}
  \includegraphics[width=0.45\textwidth]{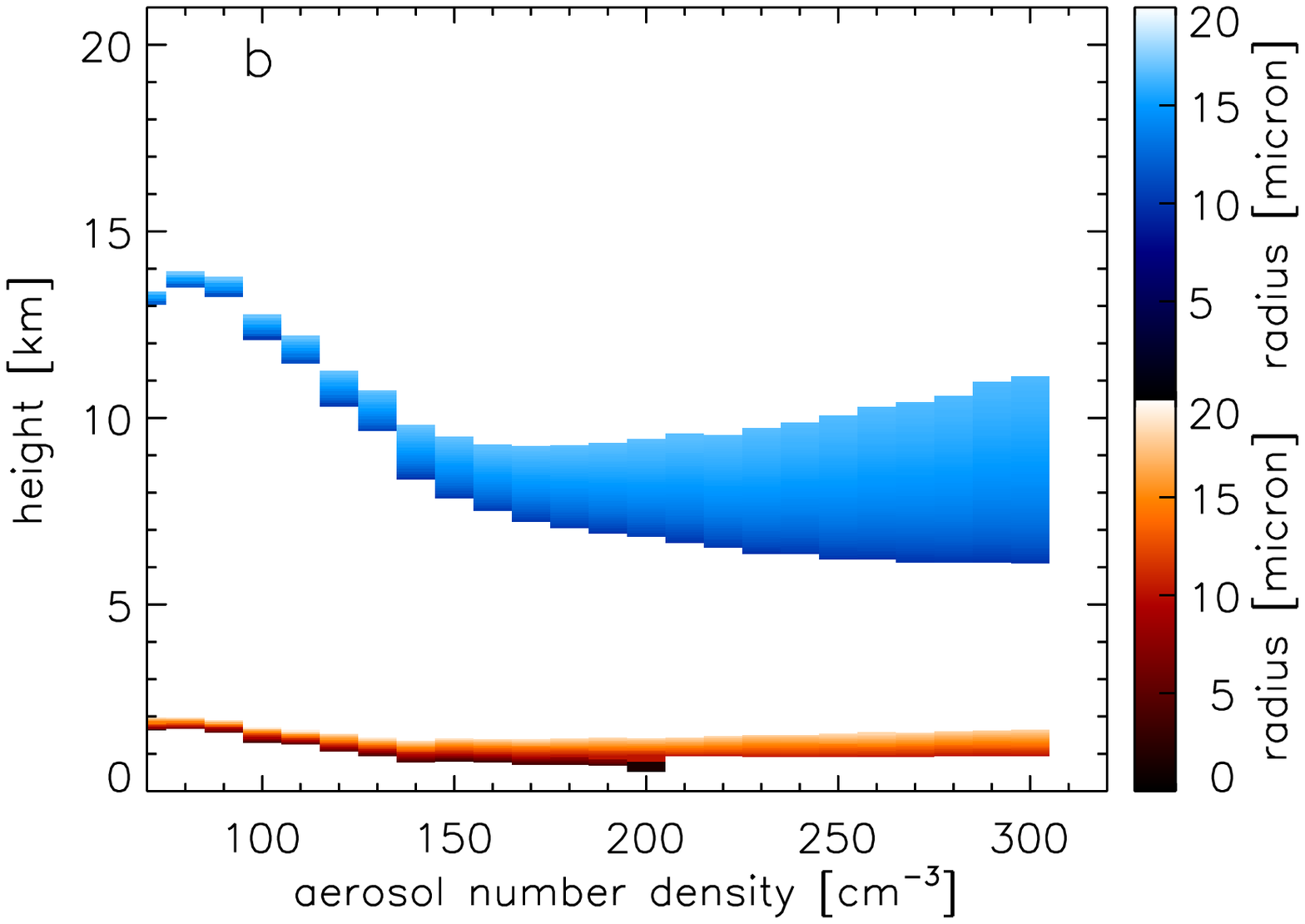}
  \caption{The number density of aerosols is varied between 70 cm$^{-3}$ and 300 cm$^{-3}$ for $f_l= 40\%$, $f_i=25\%$, relative humidity of 77\%, and precipitation efficiency of 0.8. The albedo and surface temperature are shown on the left, and the cloud properties on the right.}
  \label{fig:res_nd}
\end{figure*}

\begin{figure*}
\centering
  \includegraphics[width=0.45\textwidth]{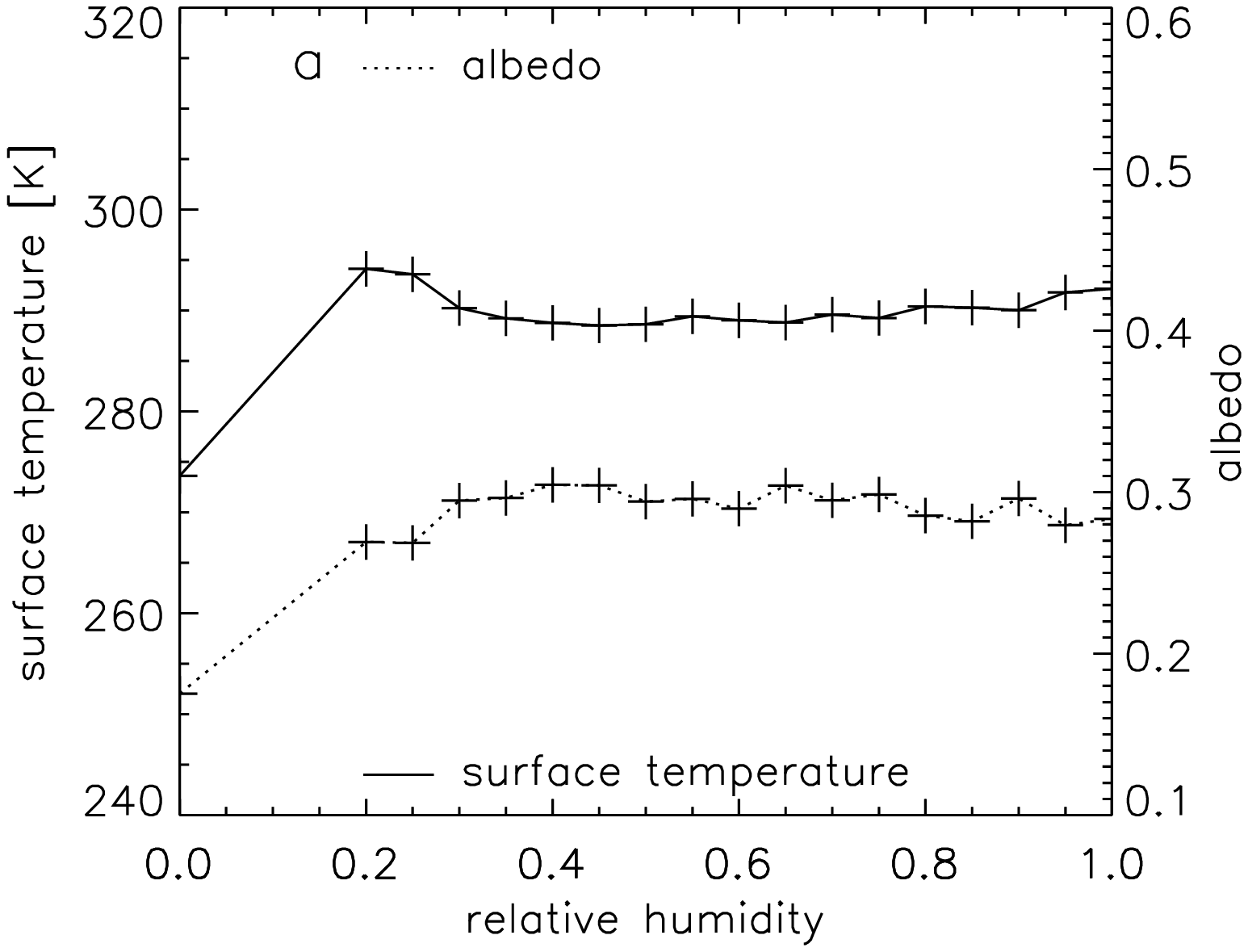}
  \includegraphics[width=0.45\textwidth]{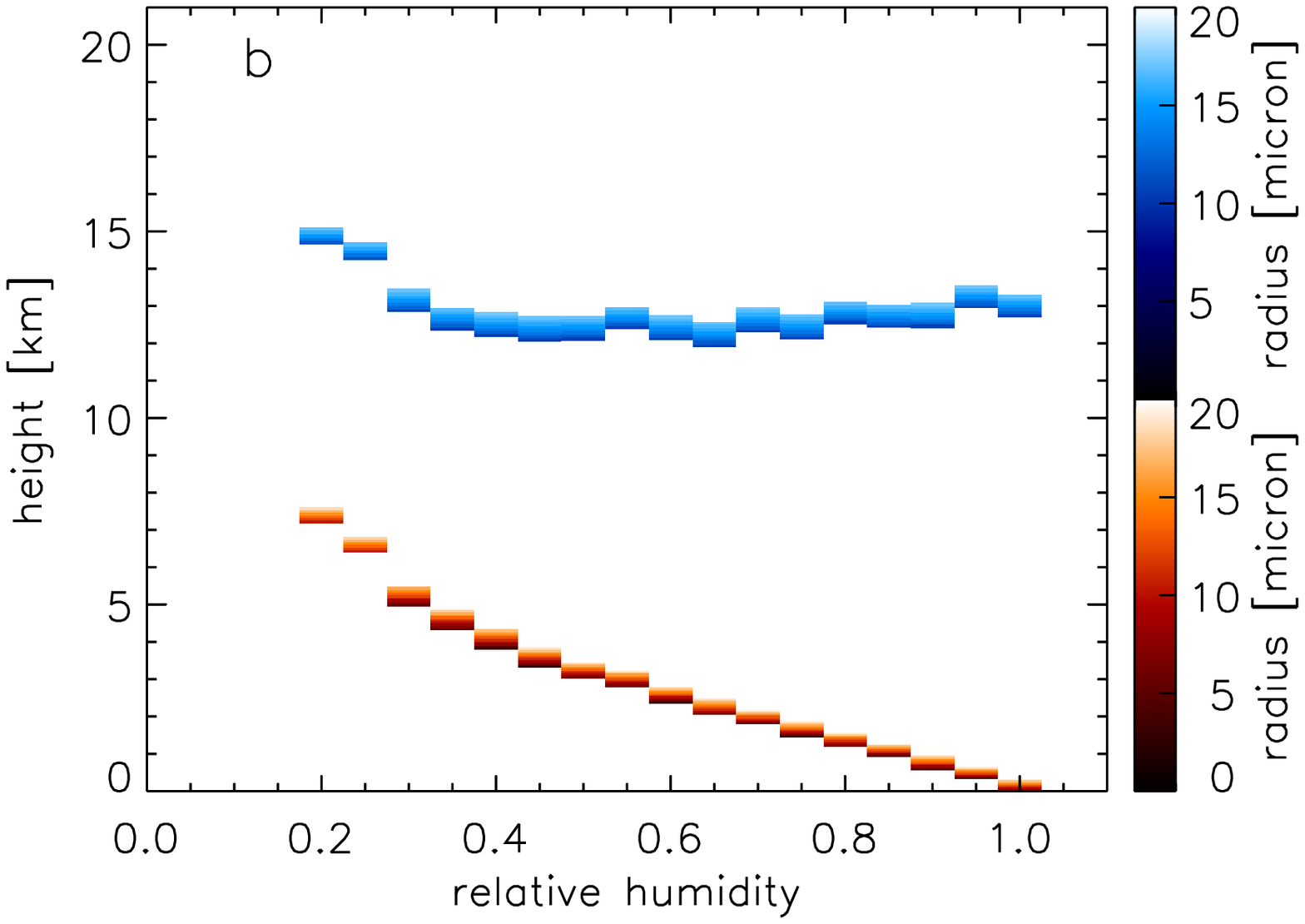}
  \caption{The relative humidity is varied between 0 and 100\% for a liquid water cloud fraction of 40\%, cirrus cloud fraction of 25\%, aerosol number density of 100 cm$^{-3}$, and precipitation efficiency of 0.8. The albedo and surface temperature are shown on the left, and the cloud properties on the right. The y axis range differs from Figs. \ref{fig:res_cfl}b and \ref{fig:res_nd}b. The dotted lines indicate the range values used on those figures.}
  \label{fig:res_relhum}
\end{figure*}

\subsection{Cloud fractions}
\label{sec:cf}
Clouds have two competing climatic effects. They reflect the incoming solar radiation thus cool the surface, and they scatter as well as absorb and re-emit the outgoing IR radiation. Whether the net climatic effect of a cloud layer is positive or negative depends on the droplet material and size, as well as the temperature at the cloud's location. If the cloud layer is located at high altitudes, thus the cloud temperature is low compared to the surface temperature, the cloud has a net warming effect on the surface. Low altitude clouds with temperatures similar to the surface temperature tend to cool the surface. These considerations are verified by our model.

We use the measured aerosols number density $n_{\mathrm{surf}}=100$ cm$^{-3}$, relative humidity ($\chi_{\mathrm{surf}}=0.77$) of Earth, set the precipitation efficiency to $e_p=0.8$ as derived in the previous section, set $f_i$ to be 0.25, and vary the liquid water cloud fraction ($f_l$) between 0 and 1 with a step of 0.05. 

Figure \ref{fig:res_cfl}a shows the albedo and surface temperature as a function of cloud fraction. Even when the liquid water cloud fraction is zero, there is an ice cloud layer present. As expected, the albedo increases (surface temperature decreases) with larger cloud fraction. Figure \ref{fig:res_cfl}a shows a linear correlation between the albedo and cloud fraction up-to 90\% cloud cover. The albedo and surface temperature for 100\% cloud coverage are 0.57 and 243 K, respectively. We also see that the surface temperature and albedo levels off above a cloud fraction of 0.9. This behavior is explained in Fig. \ref{fig:res_cfl}b, which shows the cloud properties, height of the cloudy layers and droplet radius. As the surface temperature decreases, the height of especially the ice cloud is reduced. As the temperature difference between the surface and at the location of the ice cloud (230 K) decreases, the warming effect of the ice layer vanishes, thus the surface temperature is drastically reduced. Above a cloud fraction of 90\%, the two cloud layers form one single layer and the albedo is only slightly raised by further increasing the cloud fraction.\\

In the following, we set $f_l$ to 0.4 and vary the cirrus cloud fraction between 0 and 1 with a step of 0.05. Initially, the surface temperature increases with higher cloud fraction and warms the surface (see Fig. \ref{fig:res_cfi}a). As a result, cirrus clouds form at larger altitudes, as their height is fixed at 230 K in this model (see Fig. \ref{fig:res_cfi}b). When clouds form at lower pressures, the number density of ICNs and cloud particles decrease in accordance to Eq. \ref{eq:numdens} and the clouds become optically thinner. As a result, the warming effect of cirrus clouds levels off above $f_i > 0.5$.

As seen on Fig. \ref{fig:res_cfi}b, cirrus clouds form at a height of 10-20 km. The pressure at 20 km is $7\times 10^{-2}$ bar, thus the initial number density of aerosols at this height is 7 cm$^{-3}$ according to Eq. \ref{eq:numdens}. This value is further reduced to 1.4 cm$^{-3}$ for a precipitation efficiency of 0.8. The number density of particles for a cirrus cloud at a height of 10 km is roughly four times higher. Thus cirrus clouds at a height of 20 km are thin and have a low greenhouse effect.

\subsection{Number density of aerosols}
In this section, we set the $f_l$ to 0.4, $f_i$ to 0.25, the relative humidity to 0.77, the precipitation efficiency to 0.8, and change the number density of aerosols between 70 cm$^{-3}$ and 300 cm$^{-3}$ with a step of 10 cm$^{-3}$. The number density of 70 cm$^{-3}$ is measured at oceanic areas of Earth, while a number density of 300 cm$^{-3}$ is measured in heavily populated industrial areas \citep{Miles2000}.

Figure \ref{fig:res_nd}a shows that the albedo and the surface temperature are also sensitive to the aerosol number density. If the number density of aerosols was zero, there would be no clouds on the planet, and the surface temperature would be 295 K with an albedo of 0.15, the clear sky case. We find that high aerosol number density produces geometrically (and optically) thick clouds, thus increasing the albedo and reducing the surface temperature of the planet. If the number density of aerosols is high, the same amount of water vapor can condense onto more aerosols, thus the radius of cloud droplets gradually rises with height and the cloud layer is thick (see Fig. \ref{fig:res_nd}b). If the number density of aerosols is low, the cloud layer becomes thin and the droplet size steeply increase with height. 

\subsection{Relative humidity}
\label{sec:relhum}
We set the number density of aerosols to 100 cm$^{-3}$, the $f_l$ to 0.4, $f_i$ to 0.25, the precipitation efficiency to 0.8, and vary the surface relative humidity between 0 and 1 with a step of 0.05. This way we consider the full range between desert and ocean planets. 

Figure \ref{fig:res_relhum}a shows that a relative humidity of 0\% cools the planet to 273 K. In this case, no clouds are present in the atmosphere. The surface temperature drops because an efficient greenhouse gas, water vapor, is removed from the atmosphere. The simulation with 5\%, 10\% and 15\% relative humidity are excluded because the cumulus clouds form at altitudes larger than the cirrus clouds, which is unphysical. Therefore, our cloud model is not applicable for low relative humidities. 

The climate is insensitive to changes in the relative humidity compared to the other cloud parameters. As the convective water cloud is located where an ascending air parcel becomes saturated with water vapor, higher relative humidity decreases this height. If the relative humidity is small, clouds form at high altitudes (see Fig. \ref{fig:res_relhum}b) and their green house effect dominates over their albedo effect compensating for the lack of water vapor in the atmosphere. At high relative humidities, the convective clouds form close to the surface and thus have a reduced greenhouse effect. However, the atmosphere contains large amount of water vapor to compensate. If the relative humidity is 100\%, convective clouds form already at ground level, as any vertical displacement results in super-saturation.

\section{Discussion}
\label{sec:disc}
\subsection{Critical view on the cloud model}
\label{sec:crit}

Often convection is treated as a 1D diffusion process in cloud models for brown dwarfs and giant planets \citep[see e.g.,][]{Ackerman2001}. The formed cloud droplets can be displaced to both directions and this process artificially increases the vertical extent of the cloud layer. On Earth, we see that the regions where convection is upward and downward are physically separated, thus the downward motion does not influence the cloud properties. Therefore we adopted this concept in our microphysical model.

Cirrus clouds are produced by overrunning and undercutting on Earth (Sect. \ref{sec:basic}). However, such horizontal motions cannot be self-consistently modeled in 1D codes, thus we are left to parameterize the height of the cloud deck. The adopted parametrization is probably the largest uncertainty in our model because it is solely based on Earth observations. The temperature or super-saturation level where cirrus clouds form could change, if the relative humidity, the circulation pattern of the atmosphere, etc changes. Unfortunately we are not aware of any cirrus model which could take such effects into account in a 1D model.

We consider cloud formation only by condensation, and consider one representative droplet size at a given height because the optical properties of a droplet size distribution can be approximated by a single droplet size \citep{Hu1993}. We do not simulate coalescence and the vertical motion of a droplet population. The effects of such microphysical processes are included in a simplified way in the precipitation efficiency. This parameter could in principle be constrained self-consistently by simulating the vertical motion, condensation, and coalescence of a droplet population. However, such calculation would introduce more free parameters (e.g., the distribution and material of aerosols) and add a significant computational overhead to our numerical methods. To keep the free parameters limited and the computational time short, we introduce the precipitation efficiency.

The number density of aerosols is determined by various processes such as bubble bursts on the sea, wind lifting up dust from the continents, volcanos, etc. (see Sect. \ref{sec:basic}). The total number of cloudy layers and the cloud fraction per cloud layer are determined by the circulation of the atmosphere. For example the rotation period, the friction between the surface and air, topography, relative humidity, number density of aerosols all influence the cloud fraction. The relative humidity is mostly determined by the topography of the planet, by the zonally averaged land/sea distribution and surface temperature profile. Detailed two dimensional cloud resolving models show that the precipitation efficiency depends on the water vapor content of the atmosphere, surface evaporation rate, amount of precipitating particles, zonal and vertical wind speeds, etc. \citep{Sui2007, Gao2011}. As we either do not know the above described planet properties or cannot simulate the physical processes for exoplanets, we varied these parameters independently in Sect. \ref{sec:res}. GCMs are able to resolve some of the correlations mentioned above, but such models are currently too computationally expensive to be used for parameter exploration. Therefore we were left to treat these parameters independently for now.

We use the critical Reynolds number in both cloud layers to limit the width of the cloud layers. A value of 200 is chosen for both layers, but other values could also be adopted. The critical Reynolds number determines the width of the cloud layer. If $Re$ is smaller(larger) than 200, the cloud layers become thiner(thicker) and a smaller(larger) value of precipitation efficiency would also reproduce the global average temperature and albedo of Earth. However, the critical Reynolds number should not be smaller than 70, because the flow around the droplets is laminar  and coalescence is not initiated in that case \citep{Rossow1978}. Adopting too large values of $Re$ makes the cloud too thick geometrically and optically, and a precipitation efficiency of almost unity is necessary to reproduce the observed properties of Earth. We verified by simulations that the results are not sensitive to the exact value of $Re$. Varying the other cloud parameters yield to large trends, and the effect of different $Re$ is negligible compared to these trends. 

We do not model the ice albedo feedback. The surface albedo decreases for higher surface temperature as the size of ice-covered areas become smaller due to melting. If the surface temperature decreases, the size of ice covered areas would increase, which in turn would further decrease the surface temperature. This is a positive feedback, as some cooling(warming) promotes further cooling(warming) of the surface. However, we keep the surface albedo fixed in all runs.



All our model runs were done with a fixed composition of non-condensable gases and a surface pressure of 1 bar. Comparison with Venus, Mars, and Titan indicates that the composition and mass of terrestrial planet atmospheres varies widely. Similar is expected for exoplanets. For Earth, the composition \citep{Sagan1972, Owen1979} and possibly the mass \citep{Goldblatt2009} of the atmosphere have evolved with the geochemical evolution of the planet. There are geochemical feedbacks between temperature and composition. Most famous is the temperature dependence of CO$_2$ concentration by silicate weathering, which is thought to exert a negative feedback on the global temperature on $10^5$ yrs timescales \citep{Walker1981}. Until we have high resolution spectra for terrestrial exoplanets, the atmospheric composition will be difficult to constrain -- and modeling such effects is thus beyond the scope of our climate model.

\subsection{Comparison of microphysical model to parameterized cloud models}
\label{sec:disc_cf}
Our standard simulation is described in Sect. \ref{sec:preceff} with a precipitation efficiency of 0.8. We compare this simulation to the standard simulations of \cite{Goldblatt2011} and \cite{Kitzmann2010}. Both of these cloud models are parameterized and have been applied to Earth as a test case.

\cite{Kitzmann2010} uses two cloud layers, like our model. Their model reproduces the global average temperature of Earth, but the planetary albedo is somewhat smaller (0.27) than the observed value of 0.3. The planetary albedo is calculated as the ratio of incoming to outgoing solar flux in the GEB. This implies that the outgoing solar flux of \cite{Kitzmann2010} is $10\%$ smaller than the observed value. \cite{Kitzmann2010} adopts 52 vertical grid cells and their cloud layer does not entirely fill one grid cell. The thickness of the grid cell containing the low level water cloud is for example 0.9 km, however the thickness of the cloud layer is 0.15 km (personal communication). \cite{Kitzmann2010} adopts a droplet size distribution, number density, and cloud optical depth in accordance to observations. The cloud thickness is then derived from these quantities. The grid cell containing the cloud is determined by the observed pressure at the cloud top.

The cloud heights (or pressures) of \cite{Goldblatt2011} are derived from observations. They adopt 30 vertical layers mostly centered around the troposphere and use three clouds layers. The low level liquid water cloud is resolved by 3 grid cells, the middle layer by 2, and the cirrus cloud layer by 1 grid cell. Their GEB values are shown in Fig. 5 of their paper and are in good agreement with the observed values. The largest relative error is $5\%$ in their GEB fluxes with a typical error of 2-3$\%$. 

\cite{Kitzmann2010} and our work adopt an initial droplet number density in agreement with observations. Our droplet number density is then reduced due to precipitation. The number density of droplets in \cite{Goldblatt2011} is calculated using the liquid water path, average droplet size, the height of the cloud deck and top. Their lower liquid water cloud has for example 7 droplets per cm$^3$, which is lower than in the other two models. This difference is explained by the large vertical extent of their clouds (1.4 km, compared to 0.4 km in our standard model and 0.15 km in \cite{Kitzmann2010}).

Finally, the results of parameterized cloud models differ in details from the results of the microphysical cloud model but all three models reproduce the GEB of Earth much better than clear sky models with an increased surface albedo. The advantage of our microphysical cloud model over parameterized models is that the cloud properties are determined self-consistently (except for the cirrus height), thus it is applicable to a much wider range of atmospheric and planetary properties than a parameterized model, which is tuned to Earth. The cloud model predict how cloud properties change and influence the climate, if relative humidity, cloud fraction, aerosol number density, atmospheric composition, etc is different than what is observed on Earth. Parameterized models have difficulty to cope with such issues.

\subsection{Sensitivity of the climate}
The simulations performed in Sect. \ref{sec:res} map how the structure of the atmosphere is influenced by cloud fraction, aerosol number density, relative humidity and precipitation efficiency. For precipitation efficiency between 0 and 100\%, the albedo and surface temperature varies between 0.47-0.15, and 259 K - 230 K, respectively. For a liquid water cloud fraction between 0 and 100\%, the albedo and surface temperature varies between 0.16 - 0.58, and 305 K - 242 K, respectively. For an ice cloud fraction of 0 and 100\%, the albedo and surface temperature varies between 0.26 - 0.3, and 282 K - 300 K, respectively. For aerosol number density between 70 cm$^{3}$ and 300 cm$^{-3}$, the albedo and surface temperature changes between 0.25 - 0.47, and 293 K - 262 K, respectively. The climate is least sensitive to the relative humidity. For relative humidity between 20\% and 100\%, the albedo and surface temperatures are 0.3 - 0.27, and 287 K - 294 K, respectively. A cloud-free atmosphere with zero relative humidity has a surface temperature of 273 K, and albedo of 0.17.

However, the relative humidity, cloud fraction and precipitation efficiency are not independent parameters. The relative humidity determines the amount of water vapor available for condensation and cloud formation. Therefore, assuming a zero cloud fraction and 77\% relative humidity in Sect. \ref{sec:cf} is probably unrealistic, if CCNs exist on the planet. Assuming less than 20\% relative humidity and 40\% cloud fraction in Sect. \ref{sec:relhum}, and a precipitation efficiency of both zero and unity in Sect. \ref{sec:preceff} are unlikely. The relative humidity and cloud fraction are dependent parameters, however the dependency is difficult to constrain even with global circulation models. Therefore we treat them as independent parameters. Determining the precipitation efficiency self-consistently would add new free parameters and significant computational overhead to the calculation (see Sect. \ref{sec:crit}).

It shows the intricacy of the field that the albedo and climate depends rather strongly on the aerosols. It is also observed on Earth that the cloud properties are different above land, where the aerosol number density is high, and above sea, where the aerosol number density is low \citep{Rossow1999}. Models to calculate the aerosol number density exists for Earth \citep{Seinfeld2006} and it might be beneficial in the future to generalize those models for exoplanets.

\section{Summary}
\label{sec:sum}
We developed a microphysical cloud model that can self-consistently calculate the cloud heights and droplet sizes in 1D atmosphere models. Our treatment of low level water clouds is inspired by vertical convection where warm and humid air rises from the surface and forms clouds, and dry air descends at some other location. The height of water ice clouds (cirrus clouds) is determined by a parameterization in temperature. The free parameters of the model are the surface relative humidity, number density of cloud condensation nuclei, cloud fractions, precipitation efficiency (and the temperature where cirrus clouds form, if Earth is considered).

We apply our model to Earth. As a first step, we adopted the observed relative humidity, aerosol number density, and cloud fraction values to determine what precipitation efficiency is necessary in the framework of our model to reproduce the observed average surface temperature, albedo, and GEB of Earth. We find that a 80\% precipitation efficiency is necessary to reproduce the measured GEB with good accuracy. Thus both the parameterized cloud models and our cloud model can reproduce the globally averaged observed properties of Earth well. The \cite{Goldblatt2011} model reproduces the GEB with a relative error of 2-3\%, our model with 3-4\%, and the \cite{Kitzmann2010} model has a relative error of 10\% on the visible flux leaving the atmosphere.

The advantage of our model over the parameterized models is that we calculate the cloud properties self-consistently (expect for the height of the cirrus layer), thus we can explore how the cloud properties and the climate is influenced by varying the cloud fractions, relative humidity, aerosol number density, atmospheric composition, etc. A parameterized model is tuned for an exact Earth-analog and it is challenging to extend it to a wider parameter space.

We explore how the free parameters of the model affect the albedo and the climate of the planet. The planetary climate on Earth is most sensitive to variations in the liquid water cloud fraction. The albedo varies between 0.16 and 0.58 for this parameter. The albedo varies between 0.47 and 0.15, for precipitation efficiency between 0 and 100\%. For number density between 70 and 300 cm$^{-3}$, the albedo varies between 0.25 and 0.47. The albedo varies between 0.26 and 0.3, for an ice cloud fraction between 0 and 100\%. The climate is least sensitive to changes in the relative humidity, unless it is below 20\%. The albedo is between 0.3 and 0.27, for relative humidity between 20\% and 100\%. This effect is explained by the enhanced green house effect of high altitude clouds. 

Our microphysical cloud model provides a new and valid approach for modeling clouds and their effect on the climate for a wide set of rocky exoplanets within the habitable zone and will be used to study planetary science questions related to Earth.

\section*{Acknowledgements} A. Zs. thanks Daniel Kitzmann for useful discussions regarding his cloud model. The authors thank the referees for comments that improved the manuscript.

\end{document}